\documentclass{aa}  
\usepackage{graphicx}
\usepackage{txfonts}
\usepackage{xcolor}

\begin{document}

   \title{The hierarchical formation of 30\,Doradus as seen by JWST\thanks{Based on observations with the NASA/ESA {\em Hubble Space Telescope} and the NASA/ESA/CSA {\em James Webb Space Telescope}, which are operated by AURA, Inc., under NASA contracts NAS5-26555 and NAS 5-03127.}}

   \subtitle{}
   \titlerunning{Hierarchical formation of 30\,Doradus and R\,136}

   \author{Katja Fahrion\thanks{ESA research fellow}\fnmsep
          \inst{1}
          \and
          Guido De Marchi
          \inst{1}
    }

   \institute{European Space Agency (ESA), European Space Research and Technology Centre (ESTEC), Keplerlaan 1, 2201 AZ Noordwijk, the Netherlands.
             \email{katja.fahrion@esa.int, gdemarchi@esa.int}
             }

   \date{\today}
 
  \abstract
   {The 30\,Doradus region in the Large Magellanic Cloud (LMC) is the most energetic star-forming region in the Local Group. It is powered by the feedback from the massive stars in R\,136, the 1 $-$ 2 Myr old central massive cluster. 30\,Doradus has therefore long been regarded as a laboratory for studying star and star cluster formation under conditions reminiscent of the early Universe.
    We use JWST NIRCam observations to analyse how star formation proceeds in the region. Using selections based on theoretical isochrones on colour-magnitude diagrams, we identify populations of different ages. We select pre-main-sequence (PMS) stars and young stellar objects that show excess emission from warm dust or emission lines. Studying the spatial distribution of the different populations, we find that the youngest PMS stars with ages < 0.5 Myr are located in an elongated structure that stretches towards the north-east from the central cluster.
    The same structure is found in the sources that show an infrared excess, appears to be overlapping with cold molecular gas, and covers previously investigated sites of ongoing star formation.
   Pre-main-sequence stars with ages between 1 and 4 Myr and upper main-sequence stars are concentrated in the centre of R\,136, while older stars are more uniformly distributed across the field and likely belong to the LMC field population. Nonetheless, we find stars with excess emission from on dust or emission lines as far as 100 pc from the centre, indicating extended recent star formation. We interpret the elongated structure formed by the youngest PMS stars to be an indication of the still-ongoing hierarchical assembly of the R\,136 cluster. Additionally, the lower density of old PMS stars with emission due to ongoing accretion in the central region suggests that feedback from the R\,136 stars is effective in disrupting the disks of PMS stars.}

               
   \maketitle

\section{Introduction}
The 30\,Doradus region (also known as the Tarantula nebula) in the Large Magellanic Cloud (LMC) is the most luminous and most energetic star-forming region in the Local Group. In its centre lies the OB association NGC\,2070, which itself hosts the compact subcluster R\,136 at its core, a young (1 - 2 Myr; \citealt{Crowther2016, Bestenlehner2020}) star cluster that contains some of the most massive stars known \citep{Crowther2010, Brands2022}. 30\,Dor
is therefore the closest example of a massive extragalactic star burst, powered by a young massive cluster, and it is often considered as a local analogue of the extreme star and star cluster formation conditions in more distant interacting galaxies (e.g. \citealt{Bastian2013}) or at high redshift (e.g. \citealt{Bouwens2021}). 

Consequently, understanding how star formation proceeds in 30\,Dor has been the focus of many observational campaigns across the full wavelength range.
X-ray and ultraviolet studies have established the significant contribution of the R\,136 cluster to the ionising flux and mechanical feedback in the region \citep{Bestenlehner2020, Cheng2021, Crowther2022}, but early ground-based studies already suggested that star formation is ongoing beyond the central cluster (e.g. \citealt{Walborn1991, Hyland1992, Rubio1992}). 
For example, \citealt{Walborn1987} and \cite{Walborn1991} identified three O-stars that are deeply embedded in nebular knots to the north-east and west of R\,136 with spectroscopy, while \cite{Rubio1998} detected several infrared (IR)-bright sources in the nebular filaments surrounding the central cluster with deep ground-based imaging.

Optical imaging with the \textit{Hubble} Space Telescope (HST) Wide Field Planetary Camera 2 (WFPC2) revealed highly complex nebular structures with compact dust regions and extended pillars \citep{Hunter1995a, Scowen1998, Rubio1998, Walborn2002}. Complementary to ground-based IR studies, \cite{Walborn1999} presented HST Near-Infrared Camera and Multi-Object Spectrometer (NICMOS) imaging of several fields in 30\,Doradus, resolving previously identified IR-bright sources into small clusters with multiple components. \cite{Brandner2001} used the same data to identify pre-main-sequence (PMS) stars.

Using HST Wide Field Camera 3 (WFC3) observations of the 30\,Dor nebula, \cite{DeMarchi2011a, DeMarchi2017} identified more than 1150 PMS stars in a 40 $\times$ 40 pc region. Similarly, an extended distribution of upper main-sequence (UMS) stars was found by the Hubble Tarantula Treasury Program (HTTP; \citealt{Sabbi2013, Sabbi2016}), a multi-band HST imaging survey spanning a much larger region of 30\,Dor. Using these data, \cite{Ksoll2018} identified more than 16000 likely PMS stars with a machine-learning approach. These stars are distributed across a 200 $\times$ 175 pc region in loose and sometimes filamentary structures. 

Using observations in the mid- and far-IR with the \textit{Spitzer} and \textit{Herschel} space telescopes, tens of embedded young stellar objects (YSOs) have been identified based on their colours and spectral energy distributions (e.g. \citealt{Whitney2008, GruendlChu2009, Carlson2012, Walborn2013}). Recently, \citealt{Nayak2023} reanalysed data from the \textit{Spitzer} SAGE \citep{Meixner2006} and \textit{Herschel} HERITAGE \citep{Meixner2013} surveys to identify a large sample of nearly 300 candidate high-mass YSOs in 30\,Doradus, some distributed in clumps 45 pc away from R\,136. High angular resolution ALMA CO(2$-$1) observations of the region presented by \cite{Wong2022} also found a complex network of filamentary structures in cold molecular gas, but \cite{Nayak2023} found only $\sim$ 40 \% of their YSO candidates to be associated with the CO molecular gas, suggesting that a majority of the cold H$_2$ gas is not traced by CO.

Overall, observational studies of 30\,Doradus have revealed it to be a complex region organised in clumps and filaments, in which multiple episodes of star formation have occurred in the past $\sim$ 30 Myr and are still ongoing in the molecular gas around 30\,Dor \citep{Rubio1992, Walborn1999, Kalari2018, Nayak2023}.
R\,136 was found to host multiple generations of stars as young as 0.5 Myr \citep{Crowther2016, Bestenlehner2020}, whereas the Hodge\,301 cluster, located 45 pc to the north-west, is substantially older, with an age of up to 30 Myr \citep{Lortet1991, GrebelChu2000, Cignoni2016}. Using HST, \cite{DeMarchi2017} found low-mass PMS stars as old as 50 Myr to be separated spatially from R\,136, and widely distributed O and B stars found by the VLT-FLAMES Tarantula Survey (VFTS; \citealt{Evans2011}) indicate distributed star formation within the past few million years \citep{Schneider2018}. Additionally, the massive YSO candidates identified from \textit{Spitzer} and \textit{Herschel} data by \cite{Nayak2023} suggest an increase in the star formation activity in the region in the past few million years.

In this work, we build upon existing studies of the young and old stellar populations in the 30\,Doradus region using JWST NIRCam photometry of the region. While \textit{Spitzer} and \textit{Herschel} have been instrumental in identifying deeply embedded YSOs that are invisible to the optical cameras of the HST, their limited sensitivity and angular resolution restricted the analysis to short-lived YSOs with masses of $\gtrsim$ 5 M$_\sun$. On the other hand, HST-based studies in the optical enabled identification of low-mass PMS stars down to $\sim$ 0.5 M$_\sun$ that have a considerable age spread \citep{DeMarchi2017}, but miss the most embedded phases. With its high sensitivity, high angular resolution, and coverage in the infrared, the JWST now allows us study PMS stars of various ages even in highly dust-obscured regions, as shown, for example, with the JWST NIRCam observation of NGC\,346 in the Small Magellanic Cloud (SMC; \citealt{Jones2023}) or NGC\,6822 \citep{Lenkic2023, Nally2023}. In this work, we used publicly available NIRCam data of the central field from Early Release Observations (ERO programme 2729, PI: K. Pontoppidan) as well as two NIRCam fields separated from the centre of the cluster (programme 1226, PI: De Marchi) to distinguish the various stellar populations in 30\,Doradus. 

The paper is structured as follows. Section \ref{sect:data} describes the data reduction and photometry workflow, and Sect. \ref{sect:extinction_correction} describes how we corrected for extinction. Our methods of selecting different subpopulations are presented in \ref{sect:selecting_populations}. Section \ref{sect:spatial_dist} describes their spatial distributions in the 30\,Doradus region. We discuss our findings with respect to the literature in Sect. \ref{sect:discussion} and conclude in Sect. \ref{sect:conclusions}. Throughout this work, we assume a distance to the LMC and 30\,Doradus of $D$ = 51.3 kpc, corresponding to a distance module of $\mu$ = 18.55 mag \citep{Panagia1991, Panagia2005}. At this distance, 1\arcmin\, corresponds to 14.9 pc and 1\arcsec\, to 0.25 pc, respectively.

\section{Data}
\label{sect:data}

\begin{figure}
    \centering
    \includegraphics[width=0.45\textwidth]{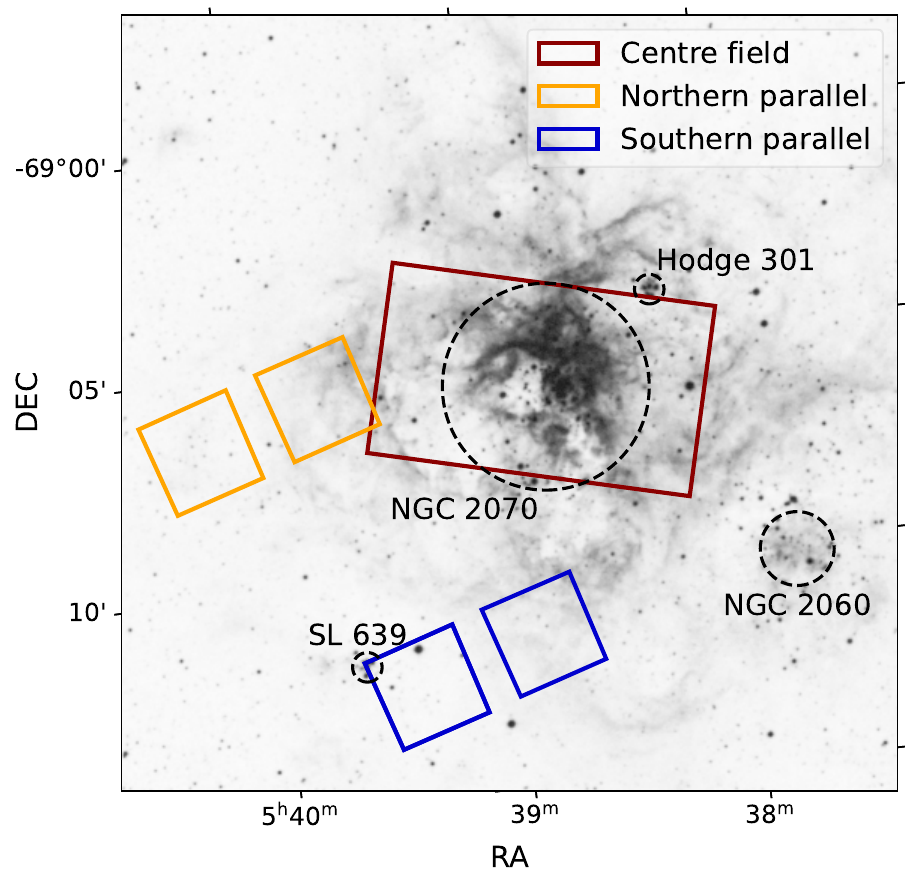}
    \caption{Greyscale DSS2 red image of 30\,Doradus. The red, orange, and blue rectangles show the footprints of the central, northern parallel, and southern parallel fields, respectively. The black circles indicate the location of well-known clusters in 30\,Doradus. North is up, and east is to the left. The greyscale image has a size of 17.5\arcmin\,$\times$ 17.5\arcmin\,(260 $\times$ 260 pc). The centre field spans a 7.4\arcmin\,$\times$ 4.4\arcmin\,region (110 $\times$ 66 pc), and the parallel observations consist of two fields each with an area of 2.2\arcmin\,$\times$ 2.2\arcmin\, (33 $\times$ 33 pc), corresponding to the footprint of the NIRCam A and B modules.}
    \label{fig:footprint}
\end{figure}

\begin{figure*}
    \centering
    \includegraphics[width=0.90\textwidth]{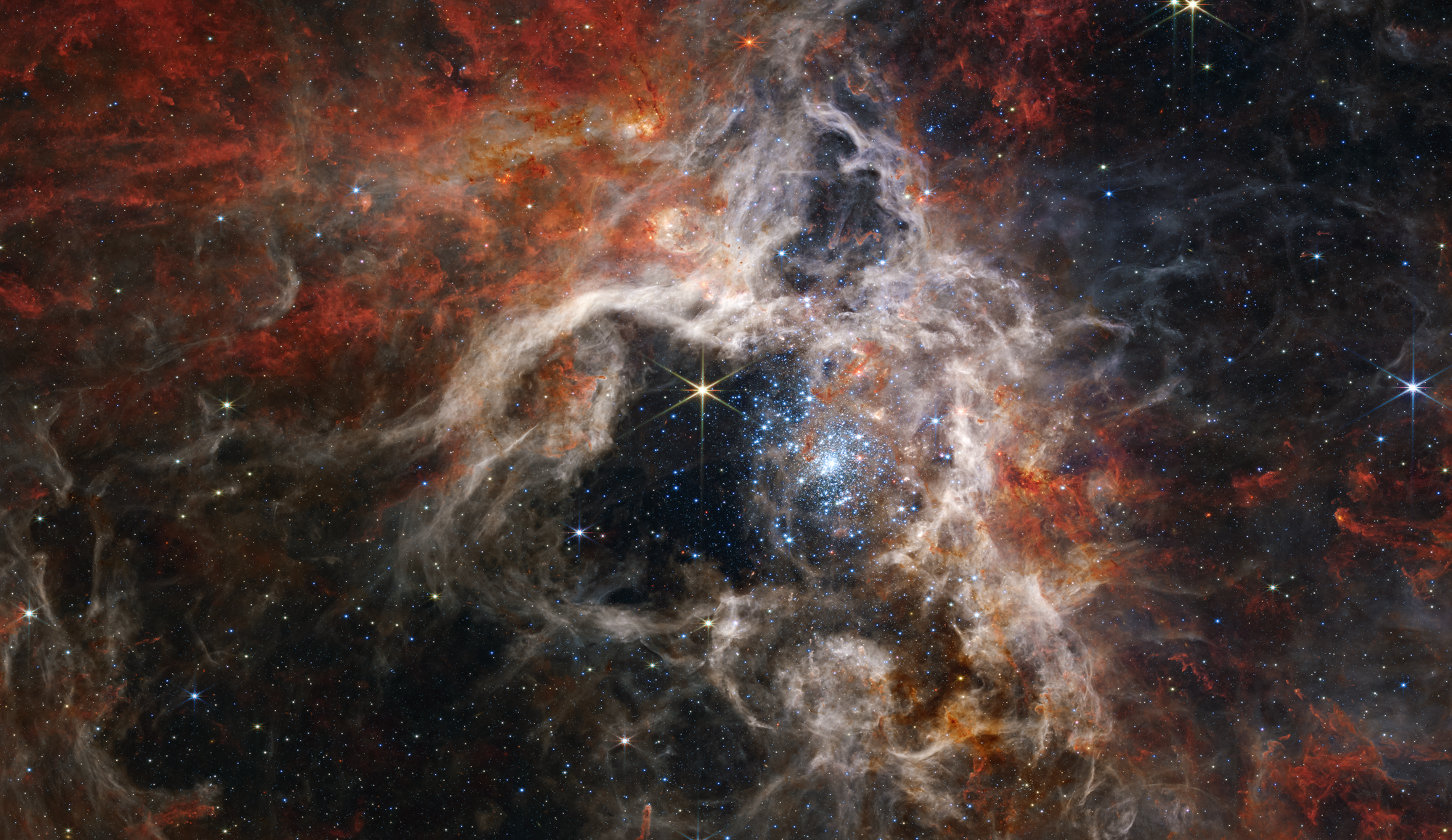}
    \caption{False-colour images of the centre field from the official image release. Red: F444W, orange: F335M, green: F200W, and blue: F090W.  The field shows a region of 7.4\arcmin $\times$ 4.4\arcmin\,(110 $\times$ 66 pc). Credit: NASA, ESA, CSA, STScI, Webb ERO Production Team.}
    \label{fig:RGB_main}
\end{figure*}

\begin{figure*}
    \centering
    \includegraphics[width=0.90\textwidth]{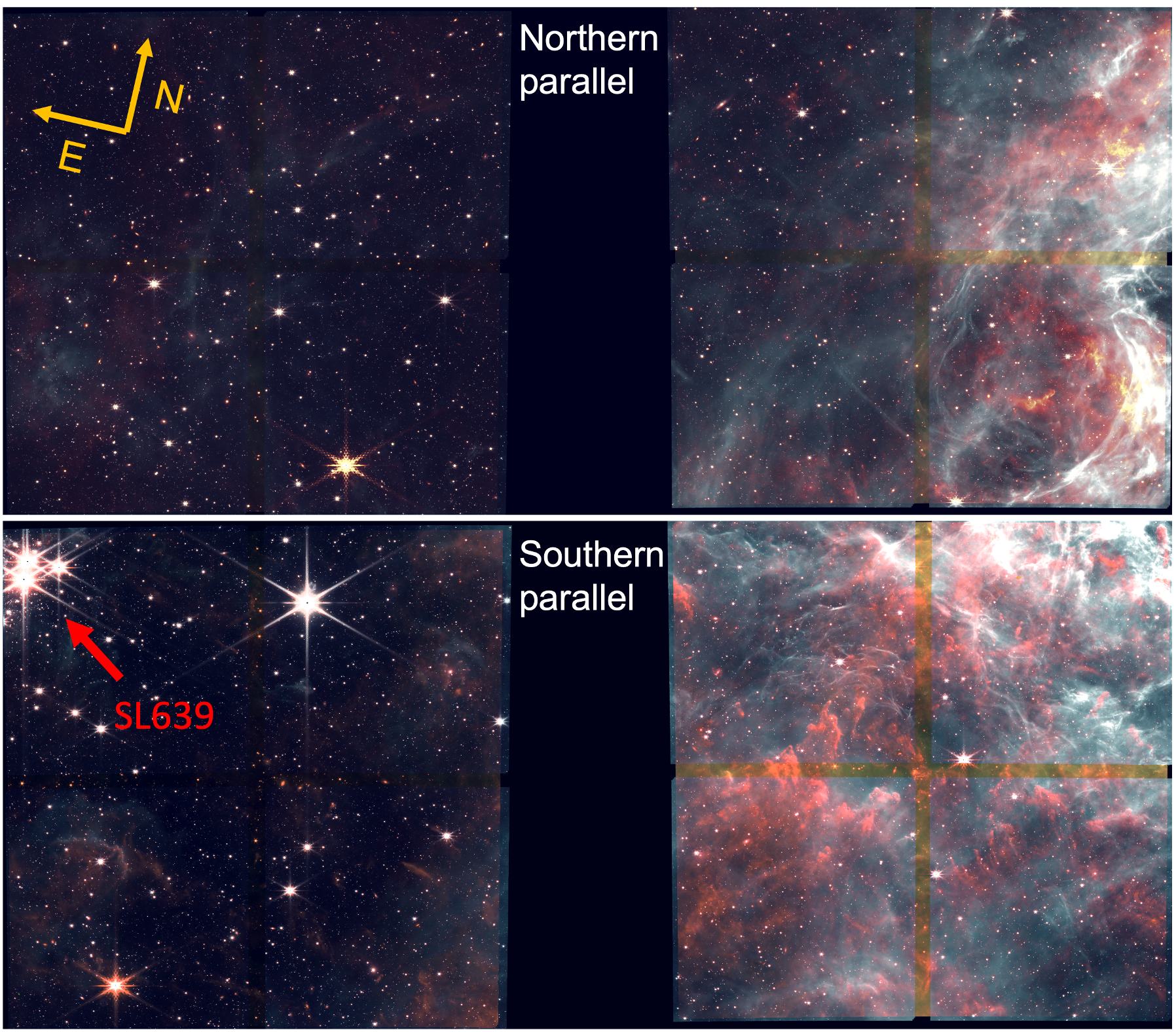}
    \caption{False-colour images of the northern parallel (\textit{top}) and southern parallel (\textit{bottom}) fields. The orange arrow indicates the orientation on the sky. The images were constructed using different JWST NIRCam filters in the respective fields. The colours for the northern parallel field are red: F430M, yellow: F405N, blue: F182M, and cyan: F187N. The colours for the southern parallel field are red: F444W, yellow: F405N, blue: F200W, and cyan: F187N. Each large square has a field of view of 2.2\arcmin\,$\times$ 2.2\arcmin\, (33 $\times$ 33 pc). The dark space in the middle corresponds to the 44$\arcsec$ gap between the NIRCam A and B modules, while the cross shape arises from the $\sim$ 5\arcsec\, gaps between the four individual detectors in the short-wavelength channel. In dithered observations such as the observations of the centre field, these gaps disappear. The location of these fields on the sky is shown in Fig. \ref{fig:footprint}.}
    \label{fig:RGB_images}
\end{figure*}

In this work, we used publicly available JWST NIRCam data from the Early Release Observations programme 02729 as also described in \cite{FahrionDeMarchi2023}, where we used the JWST data in combination with HTTP HST data to derive the extinction law in 30\,Dor from 0.3 to 4.7 $\mu$m. Figure \ref{fig:RGB_main} shows the image from the offical release\footnote{\url{https://webbtelescope.org/contents/news-releases/2022/news-2022-041}}. Here, we complemented the public JWST data of the central region of 30\,Doradus (called central field or centre field in the following) with two additional NIRCam pointings that were observed on 8 July 2023 as parallel observations to NIRSpec observations in programme 01226 (PI: De Marchi). We refer to these fields as the northern and southern parallel fields. Table \ref{tab:data} lists basic details of the observations such as the filter combinations and exposure times. Figure \ref{fig:footprint} shows an overview of the 30\,Doradus region with a DSS2 red image and the footprint of the JWST data highlighted.

\begin{table}[]
    \caption{Overview of NIRCam observations.}
    \centering
    \begin{tabular}{c c c}\hline \hline
    Field & Filter & Exposure time\\ 
        &   & (seconds) \\ \hline
    Centre & F090W & 182 \\
    Centre & F187N & 289 \\
    Centre & F200W & 193 \\
    Centre & F335M & 182 \\
    Centre & F444W & 193 \\
    Centre & F470N & 289 \\ \hline
    Northern parallel & F182M & 451 \\
    Northern parallel & F187N & 451 \\
    Northern parallel & F405N & 451 \\
    Northern parallel & F430M & 451 \\ \hline
    Southern parallel & F187N & 451 \\
    Southern parallel & F200W & 451 \\
    Southern parallel & F212N & 451 \\
    Southern parallel & F405N & 451 \\
    Southern parallel & F444W & 451 \\
    Southern parallel & F470N & 451 \\ \hline 
    \end{tabular}
         \tablefoot{Exposure times refer to individual exposures. The parallel observations include three exposures per filter. The central field is a mosaic of 20 individual pointings, and the depth is not uniform across the field of view, but most sources were covered by at least three exposures.}
    \label{tab:data}
\end{table}

\subsection{Data reduction}

For this paper, we further refined our photometric catalogues presented in \cite{FahrionDeMarchi2023}. Firstly, we reran the NIRCam data reduction pipeline steps 2 and 3 using \textsc{jwst\_pipeline} version 1.10.02 with the CDRS context map \textsc{jwst\_1089.pmap}. We aligned the world coordinate systems to the Gaia DR3 catalogues. The main difference to the older data reduction version used in \cite{FahrionDeMarchi2023} is the treatment of bad-quality pixels that were set to zero in the old version and are now NaN values. This leads to a smoother appearance in the final mosaics, but does not affect the derived photometry.

\subsection{Source detection and photometry}
While \cite{FahrionDeMarchi2023} used only the bright red clump stars, we are now interested in the full stellar populations. For this reason, we repeated the source extraction and photometry. For source detection, we used the combined mosaics with an optimised background subtraction to improve the detection of faint sources in regions with strong nebular emission. As in \cite{FahrionDeMarchi2023}, we used \textsc{photutils MMMBackground} to estimate and subtract the 2D background. Then, we detected all sources in the background-subtracted image that have two connected pixels 3$\sigma$ above the root mean squared of the background. For the narrow-band filters such as F187N and F470N, we used a 2$\sigma$ threshold. Then, this master list from the mosaics was used to obtain forced-aperture photometry on the individual exposures using an aperture radius of 2.5 pixels. The background was determined in an annulus with an inner radius of 4 and an outer radius of 5.5 pixels and was subtracted from the aperture flux.
To ensure that the world coordinate system was consistent between mosaics and individual exposures, we used calibrated but not combined \textsc{tweakregstep} files. They were corrected for the 1/f striping using the routine of Chris Willot\footnote{\url{https://github.com/chriswillott/jwst/}}.

As another advancement, we now also performed PSF photometry using \textsc{photutils} with the \text{WebbPSF} models version 1.1.0. We used a grid of 25 PSFs per detector. For the aperture photometry, the PSF photometry was applied to the individual exposures using the master catalogue from the respective mosaic, but the centroid positions were fitted freely. We note that this only leads to small adjustments on a sub-pixel level in the centroid positions and does not lead to source confusion, but optimises the fits. For each exposure, the two-dimensional background was derived using a box size of 5 $\times$ 5 pixels and a filter size of 3 pixels. This background was then supplied to the PSF-fitting routine. We also experimented with directly creating PSF models from the data. However, due to the strongly varying background, we found that the \textsc{WebbPSF} models perform better. The corresponding aperture corrections are listed in Appendix \ref{app:aperture_corrections}.

For each filter, we first obtained the photometry in each exposure and then combined this into a master table. We calculated the magnitudes as the mean from each individual exposure and the corresponding uncertainties as the standard deviations. Then, the catalogues for each filter were combined into a master catalogue, considering only sources that were detected in at least two JWST filters. The final catalogue of the centre field has 149589 sources. Out of these, 109782 (73 \%) can be matched to the HTTP catalogue in the NIRCam footprint. Of those without a HST match, more than half are redder than F090W $-$ F200W = 1.5 mag.

For the parallel observations, we used the same data reduction and source extraction steps. In the northern parallel fields, 26610 sources are found in at least two of the four filters, but only 3465 of them have a match to the HTTP because the overlapping footprints are small. In the southern parallel observations, 37680 sources are detected, 12636 of which are matched to the HTTP. Although the exposure time is the same for both parallels, the larger number of filters and the choice of wide filters over medium wide filters in the southern parallel observations leads to the larger number of detected sources. 

\subsection{Colour images of the parallel observations}
Figure \ref{fig:RGB_images} shows false-colour images of the two parallel fields, created by combining all of the available NIRCam filters. Red represents long wavelengths, and blue represents short-wavelength channels. 

These images clearly show nebulosity with a high level of detail in both parallel fields, in particular, in the western parts of the observations, which are closer to the centre of 30\,Doradus. 
The eastern parts also show faint nebular features. The north-west corner of the southern parallel field covers part of the SL\,639 cluster, for which \cite{Britavskiy2019} reported an age of \mbox{$\sim$ 22 Myr} based on VFTS data (see also \citealt{Evans2015}).

\section{Correcting for extinction}
\label{sect:extinction_correction}

\begin{figure}
    \centering
    \includegraphics[width=0.49\textwidth]{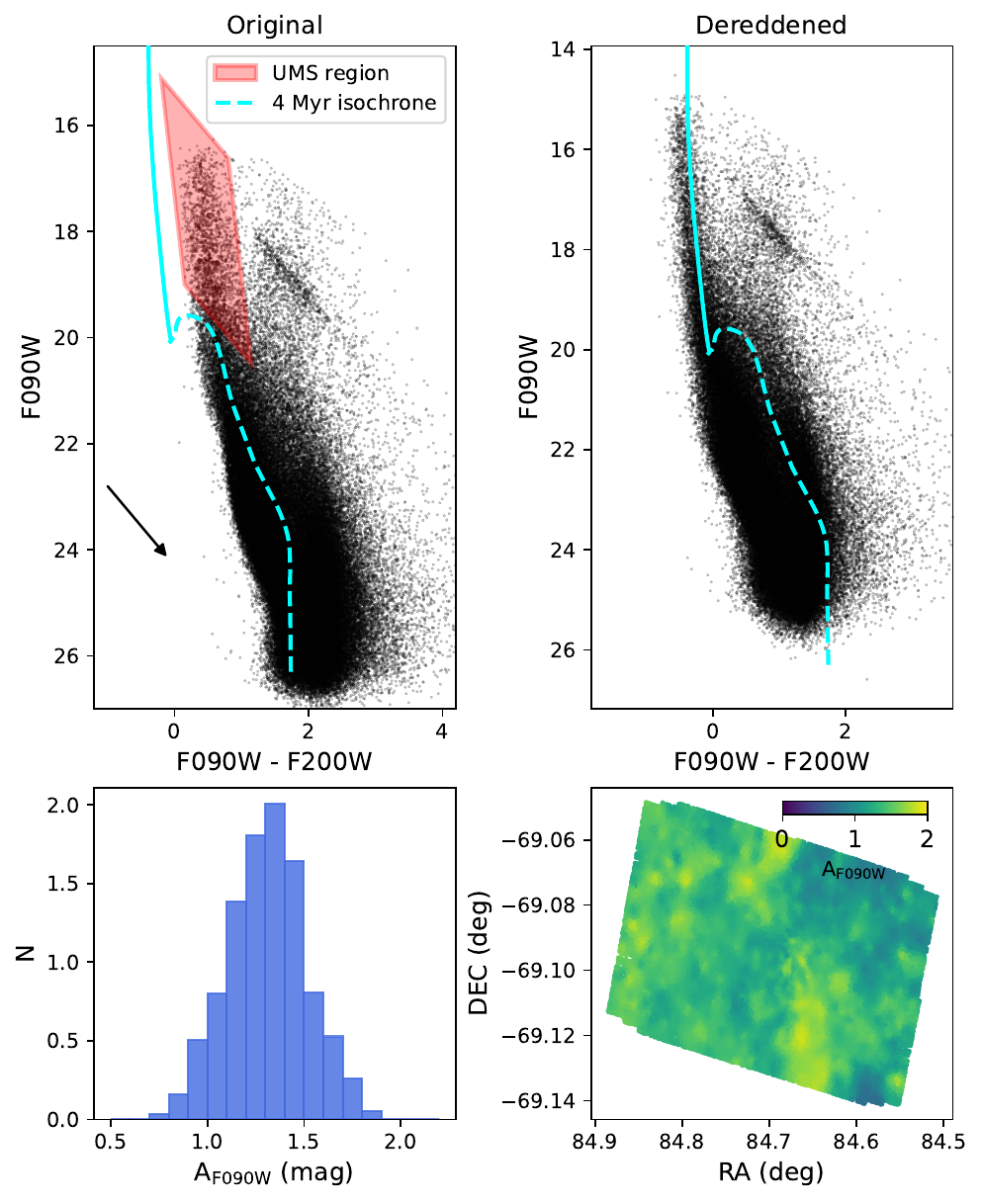}
    \caption{Extinction correction for the centre field. \textit{Top left}: Original F090W - F200W CMD. The cyan line represents a 4 Myr (log(Age) = 6.65) MIST isochrone with metallicity [Fe/H] = $-$2.5 dex (solid in the upper main sequence). The red parallelogram shows our selection of UMS stars that we used to derive the extinction by placing them back onto the isochrone. \textit{Top right}: Extinction-corrected CMD. \textit{Bottom left}: Normalised histogram of the derived extinction values A$_\text{F090W}$. The mean is at 1.31 $\pm$ 0.20 mag, and the arrow in the top left panel corresponds to this extinction. \textit{Bottom right}: Map of derived extinction values. The smooth appearance is caused by the smoothing parameter $\epsilon$.}
    \label{fig:extinction_corr}
\end{figure}

We derived the extinction law in 30\,Doradus in \cite{FahrionDeMarchi2023} using the same JWST NIRCam data in the centre field as well as archival HST data from the HTTP. We now used the derived law to correct the colour-magnitude diagrams (CMDs) of the central field for extinction using a similar approach as described in \cite{Ksoll2018} or \cite{DeMarchi2011, DeMarchi2017}. As we are mainly interested in correcting young populations and PMS stars, we based our extinction correction on UMS stars for which the absolute extinction can easily be derived by placing them back onto a theoretical isochrone. We first selected UMS in the F090W $-$ F200W CMD as shown in Fig. \ref{fig:extinction_corr} and then placed each UMS star back to its theoretical location on an unreddened 4 Myr MIST\footnote{\url{https://waps.cfa.harvard.edu/MIST/}} isochrone \citep{Dotter2016, Choi2016} with [Fe/H] = $-0.35$ dex (\citealt{Choudhury2021}, see also Sect. \ref{sect:isochrones}). This isochrone already includes also extinction in the Galactic foreground. For this component, we used the extinction law from \cite{Cardelli1989}, with $R_V = 3.1$, and $A_V = 0.22$, which is the value that is generally adopted for this line of sight (e.g. \citealt{Brunet1975, Isserstedt1975, FitzpatrickSavage1984}). As the extinction law derived in \cite{FahrionDeMarchi2023} has only considered the slope of the extinction vector rather than the absolute extinction value, we did not attempt to consider the total extinction correction, which would also include foreground extinction (see e.g. \citealt{MaizApellaniz2014} ).
In this way, we calculated the absolute extinction in F090W for all selected UMS stars across the field of view. 

We then used the UMS stars to derive the F090W extinction for the full sample of stars by calculating the distance-weighted extinction of the $n$th nearest UMS stars (see also \citealt{Ksoll2018}),
\begin{equation}
    A_{\text{F090W}} = \sum_{n=1}^{N} w_n A_{\text{F090W}_n}^{\text{UMS}},
\end{equation}
with weights $w_i$,
\begin{equation}
    w_i = \frac{1}{d_{i}^{2} + \epsilon^{2}} \frac{1}{\sum_{n=1}^{N}\frac{1}{d_{n}^{2} + \epsilon^{2}}},
\end{equation}
where $d_i$ is the distance to the $n$th nearest UMS star, and $\epsilon$ is a smoothing parameter in the same units as $d_i$. Using these parameters, we find an mean extinction of A$_\text{F090W} = 1.31 \pm 0.20$ mag, as shown in the bottom left panel of Fig. \ref{fig:extinction_corr}. 

After testing, we chose to use the 20 nearest UMS stars with a smoothing of $\epsilon$ = 5\arcsec\,around each star in the catalogue. To test this correction, we also applied it to the UMS stars, but in this case, we first removed the respective UMS star from the sample and then used other nearby UMS stars to derive its mean extinction. We found that these parameters resulted in a narrow UMS sequence that closely matched the theoretical isochrone. The right panel in Fig. \ref{fig:extinction_corr} shows the extinction-corrected CMD in F090W - F200W. We used the extinction law from \cite{FahrionDeMarchi2023} to also correct all stars in the other JWST NIRCam bands.

The extinction map in the bottom right panel of Fig. \ref{fig:extinction_corr} shows the distribution of the derived extinction values, clearly illustrating that the extinction follows the optical appearance of the nebula.  Fig. \ref{fig:extinction_corr} shows that the extinction-corrected CMD fits the \mbox{4 Myr} isochrone much better and shows a more narrow UMS. The still extended red clump shows that the correction is not perfect, especially for old populations. However, this is not expected either because by choosing to use the UMS stars for correction, we aimed to correct the extinction for the young populations. The old populations such as red giant stars are not spatially correlated with the UMS stars and therefore cannot be dereddened using UMS stars as reference. Additionally, we caution that we made the assumption that PMS stars follow the same extinction law as the UMS, even though the intrinsic colours of these populations are different and non-linearities in the extinction can have an effect, especially at large absolute extinctions (e.g. \citealt{MaizApellaniz2014, MaizApellaniz2018}). Following \cite{MaizApellaniz2020}, at absolute extinctions in F090W of $\sim$ 1.3 mag, we might expect to see effects at the level of 0.05 mag, similar to the typical uncertainty of the PMS stars. Consequently, disentangling this effect with our data is challenging, especially given that the typical scatter in extinction values of neighbouring UMS stars are about 0.1 mag.

We did not apply a similar extinction correction in the parallel fields because the filter choices were made to identify sources with emission line excess (see Sect. \ref{sect:excess_selections}). As a consequence, no colour combinations are available in which the upper main sequence can be well separated from the red giant branch. Additionally, only a small subset of stars in the parallels overlap with the HTTP footprint and have available HST colours that could be used to derive the extinction. However, we note that we are still able to identify PMS stars and YSOs based on their excess emissions, as described below. 

\section{Selecting subpopulations}
\label{sect:selecting_populations}
In the following, we describe our approaches to select subpopulations of different ages using colour-magnitude and colour-colour diagrams. For the central field, we used extinction-corrected data.

\subsection{Comparison with isochrones}
\label{sect:isochrones}
\begin{figure}
    \centering
    \includegraphics[width=0.48\textwidth]{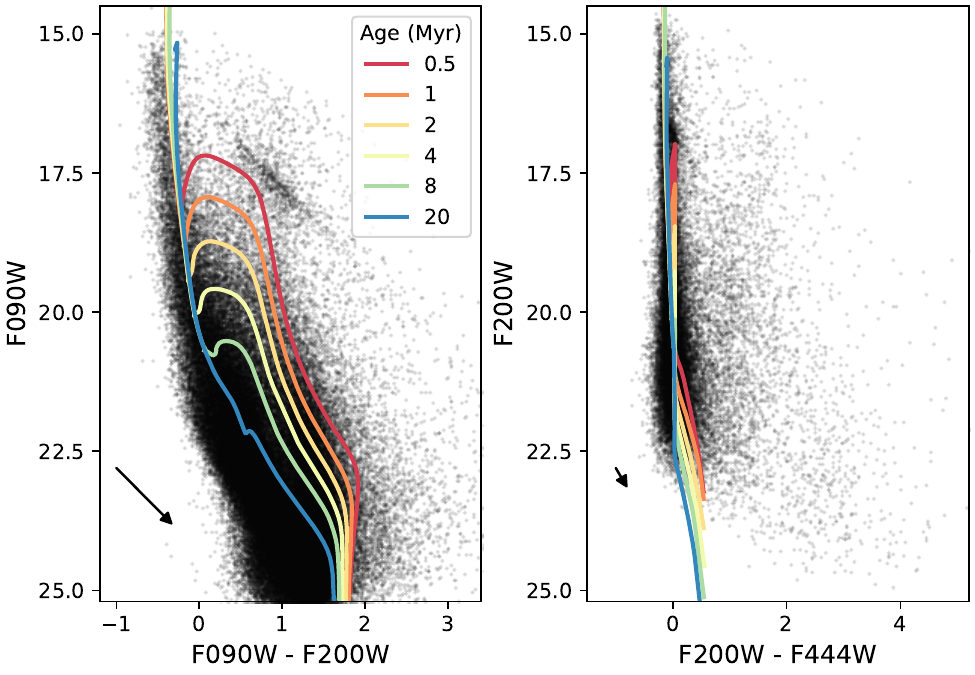}
    \caption{Extinction-corrected CMDs of the central field with MIST isochrones between 0.5 and 20 Myr. The different panels show different colour combinations. Only stars with photometric uncertainties smaller than 0.5 mag are plotted. The black arrows indicate the direction of extinction.}
    \label{fig:isochrones}
\end{figure}

We compare the extinction-corrected CMDs of the centre field in Fig. \ref{fig:isochrones} with theoretical MIST isochrones with ages of 1, 4, 8, and 20 Myr. The isochrones use [Fe/H] = $-0.35$ dex (see \citealt{Choudhury2021}) and a Milky Way foreground extinction of $A_V = 0.22$ mag following the law from \cite{Cardelli1989}.

The CMD shows that isochrones with ages even younger than 1 Myr are needed to describe the red and faint populations, but some stars are significantly redder than any of the isochrones. This is particularly clear in the right panel, which shows the F200W $-$ F444W CMD. Here, a population of sources with a significant excess towards longer wavelengths is visible. Additionally, the comparison between these two panels illustrates that two blue JWST filters with sufficient separations in their central wavelengths (e.g. F090W and F200W) are needed to different populations using isochrones. Because filters like this are missing in the parallel observations, comparisons with isochrones cannot be used there to distinguish between populations. 

\begin{figure}
    \centering
    \includegraphics[width=0.39\textwidth]{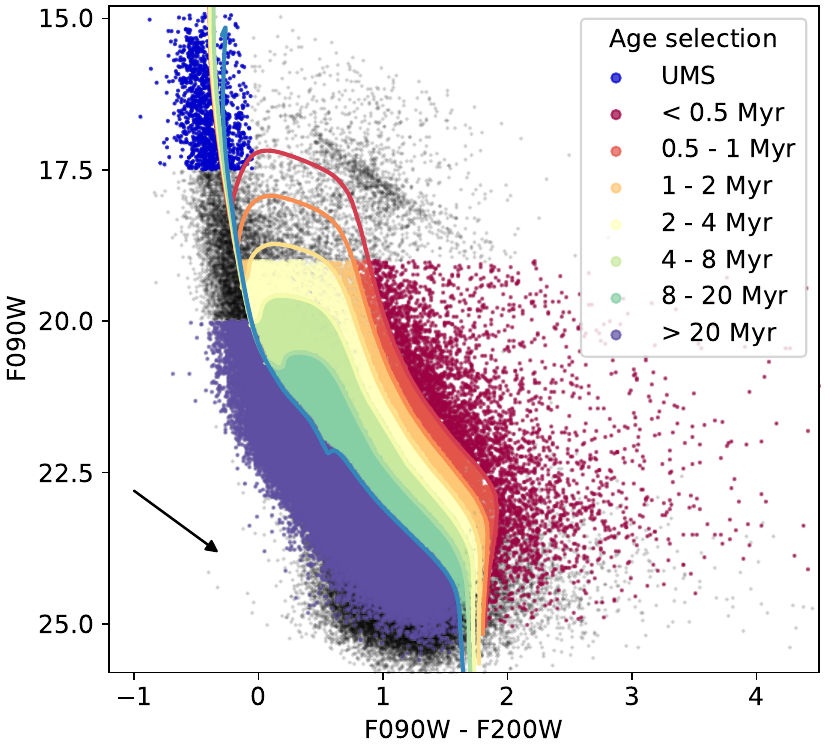}
    \caption{Selecting populations of different ages in a extinction-corrected CMD. The lines show the same isochrones as in Fig. \ref{fig:isochrones}. The coloured dots show the selected stars of different ages. We only selected stars fainter than F090W = 19.5 mag to avoid confusion with older giant stars.}
    \label{fig:isochrones_selection}
\end{figure}

We used the MIST isochrones to make a simple selection based on the expected ages of different populations in 30\,Doradus. Firstly, we again selected the UMS stars in the CMD, this time, using a simple rectangle and a magnitude limit of F090W = 18.5 mag. For the faint populations, we only considered stars fainter than F090W = 19.5 mag to avoid confusion with giant stars. In the same way as for the youngest stars, we selected all stars redder than the 0.5 Myr MIST isochrone, as shown in Fig. \ref{fig:isochrones_selection}. PMS stars with intermediate ages were selected to lie in the region bounded by the other isochrones, selecting stars with ages between 0.5 and 1, 1 and 2, 2 and 4, 4 and 8, 8 and 20, and stars older than 20 Myr. For the oldest stars, we chose an even stricter brightness cut of F090W = 20 mag. To account for photometric uncertainties, we only made these selection for stars with uncertainties \mbox{$< 0.1$ mag} in both the F090W and F200W filters. We note that this selection reaches down to F090W = 25 mag, corresponding to an initial mass of 0.5 $M_\sun$ for a 20 Myr isochrone, 0.25 $M_\sun$ for a 4 Myr isochrone, or 0.1 for a 1 Myr isochrone. This shows that low-mass PMS stars are included. At the same time, the brightness cut at 19.5 mag sets a limit to stars less massive than 3 $M_\sun$. 

As the focus of this work is to investigate how JWST NIRCam can be used to distinguish different populations, we did not consider a more refined fitting of ages and masses, as is often done with HST data (e.g. \citealt{DeMarchi2017}).
However, we note that especially when the extinction is very high, the reddening correction can be a source of uncertainty in the age determination, and in individual cases, it can move stars from one selection to the next.

\subsection{Populations with excess emission}
\label{sect:excess_selections}
\begin{figure*}
    \centering
    \includegraphics[width=0.95\textwidth]{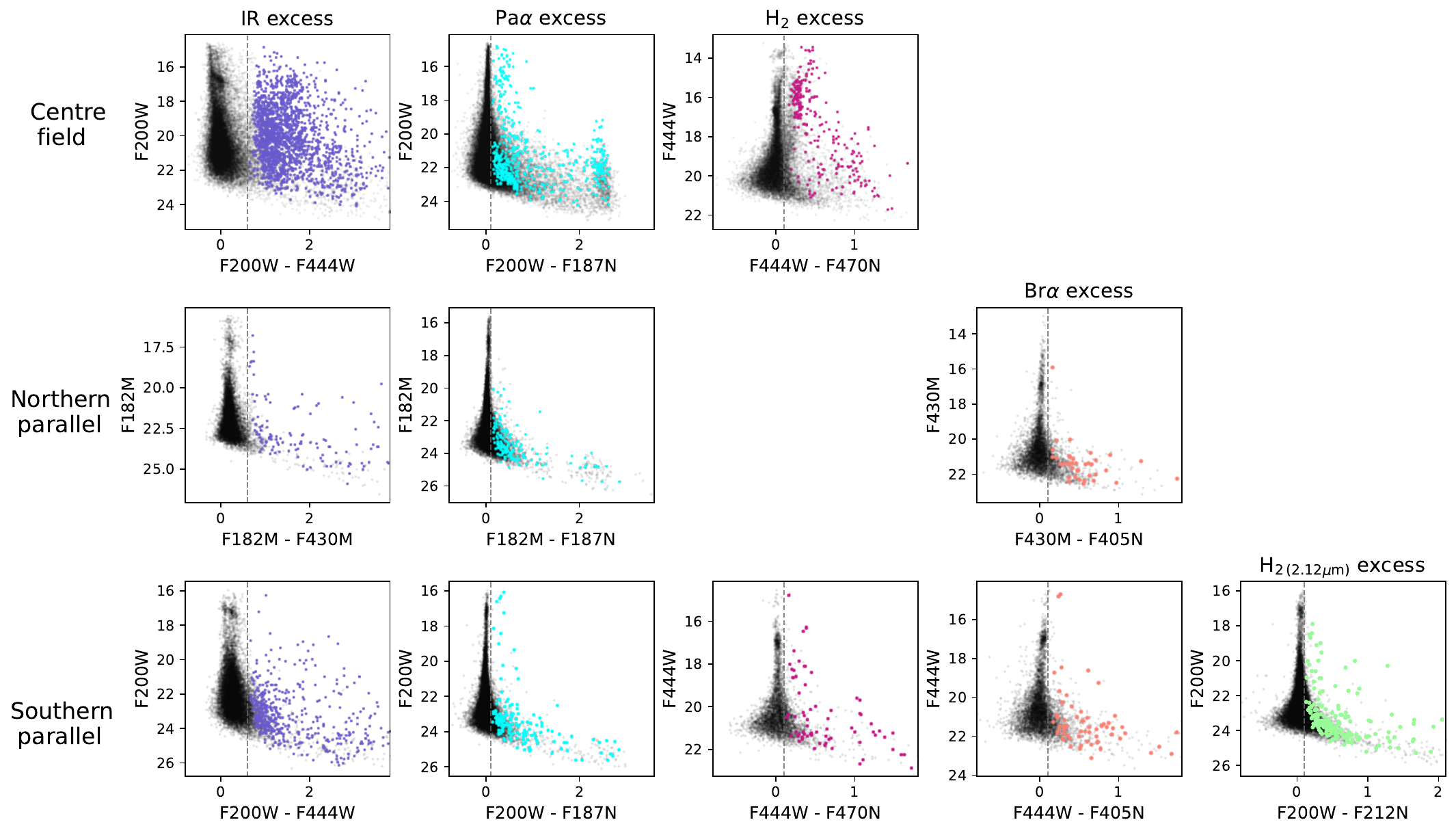}
    \caption{Selecting sources with excess emission. Each panel shows a different filter combination that was used to select sources (from left to right) with IR, Pa$\alpha$, H$_2$, Br$\alpha$, and H$_{2\,(2.12 \mu \text{m})}$ excess. The selected sources are shown with coloured dots. As a result of the filter choices, additional influence of reddening is very small.}
    \label{fig:CMDs_excess}
\end{figure*}

The large number of filter combinations that also consider multiple narrow-band filters allows us to select and study stellar populations within 30\,Doradus that show an excess compared to the main populations. This excess can either be caused by emission from certain emission lines (e.g. the HI Pa$\alpha$, Br$\alpha$, or H$_2$ lines), or from warm dust that emits at long wavelengths. 
We selected sources that show excess emission using colour cuts, as shown in Fig. \ref{fig:CMDs_excess}. This figure shows CMDs using the colours described below.

We considered excess towards the mid-infrared (called IR excess in the following) using the F200W - F444W colours (or F182M - F430M for the northern parallel field). To select excess due to emission from the Pa$\alpha$ line, we used F200W - F187N (or F182M - F187N) colours, and for excess due to emission from molecular hydrogen, called H$_2$ excess, we used F444W - F470N (for the centre field and southern parallels). Additionally, the parallel observations were covered with the Br$\alpha$ narrow-band filter F405N, so that we used F444W - F405N for the southern and F430M - F405N for the northern parallel field to select sources with excess in this emission line. Finally, the southern parallel observations also used the F212N narrow band, which covers the 2.12 $\mu$m line of H$_2$. To select excess sources, we used the F200W - F212N colour in this case. 

Our colour cuts are listed in Table \ref{tab:excess_selections}. We required the selected sources to have larger (redder) colours than these thresholds. To account for the photometric uncertainties, we only considered sources with magnitude uncertainties below 0.1 mag. Additionally, we only selected sources that met the colour cuts with at least 5$\sigma$ significance, where $\sigma$ is the combined uncertainty of the respective two filters. 

The last column of Table \ref{tab:excess_selections} reports the resulting number of stars that meet these criteria. They are plotted in different colours in the CMDs of Fig. \ref{fig:CMDs_excess}. Especially the centre field, many of the excess sources also include bright stars, whereas the parallel observations farther away from the centre of 30\,Doradus tend to show fewer excess sources and fainter sources, illustrating that especially massive star formation is mainly limited to the central region. Nevertheless, there are still sources with IR or emission line excess in the parallel fields, clearly indicating ongoing or recent star formation in the parallel fields. These sources lie as far as $\sim$100 pc from the centre of 30\,Doradus. 

We note that visual inspection of the sources with an Pa$\alpha$ excess higher than two magnitudes revealed that around half of them are extended sources in the F187N filter. These mini-HII regions are frequently not detected in F090W.

\begin{figure}
    \centering
    \includegraphics[width=0.49\textwidth]{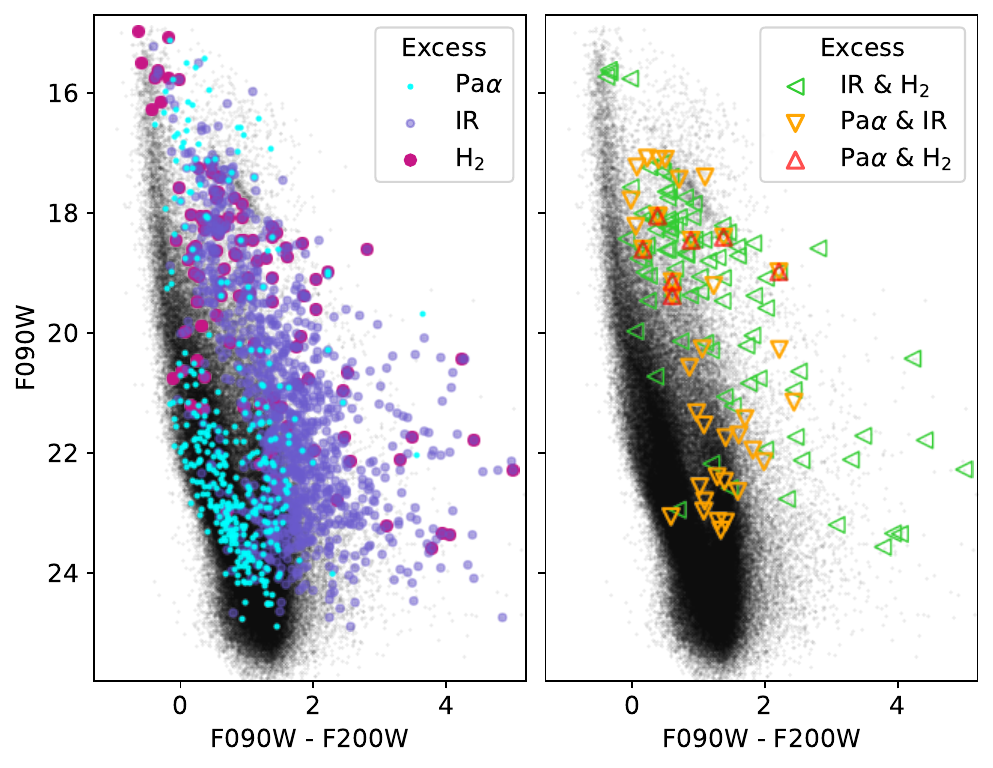}
    \caption{F090W - F200W CMD for the centre field. \textit{Left}: Different excess populations are shown by the coloured dots. \textit{Right}: Triangles show sources that show excess in two or more selections.}
    \label{fig:CMD_center_excess}
\end{figure}

A combined comparison of the different populations in the centre field is shown in Fig. \ref{fig:CMD_center_excess}. The left panel of this figure reveals that the IR-excess sources are redder than the main populations in this CMD, while the H$_2$-excess sources scatter across the full range of colours, but are mainly relatively bright due to the limited depth of the F470N observations. In some red and bright Pa$\alpha$ sources, the Pa$\alpha$ emission is probably caused by winds around massive stars (possibly Be stars), but there might also be some massive PMS stars in this selection. Stars with faint Pa$\alpha$ excess show a bimodal distribution. Some of them are close to the main sequence, while others lie towards redder colours, in the area of PMS stars. This is similar to the results of \cite{DeMarchi2017}, who used HST H$\alpha$ narrow-band observations in conjunction with broad-band filters to identify accreting PMS stars. In this way, they identified bona fide PMS stars with locations both on PMS tracks and close to the main sequence. This was interpreted to show that older PMS stars that have almost reached the main sequence can also still show an excess in hydrogen emission, likely caused by ongoing accretion. With NIRSpec spectroscopy of similar targets in NGC\,346, this emission was then confirmed. This clearly shows that PMS stars in the low-metallicity environment of the SMC can accrete for several dozen million years \citep{DeMarchi2023}.

\begin{table}[]
    \caption{Overview of selected populations with excess emission.}
    \centering
    \begin{tabular}{c c c c c} \hline \hline
     Field & Excess   &  Colour & Threshold & $N$ \\
                &  & & (mag) \\ 
        \hline
      Centre &  IR & F200W - F444W & 0.60 & 1637\\
      Centre & Pa$\alpha$ & F200W - F187N & 0.10 & 518 \\
      Centre & H$_2$ & F444W - F470N & 0.10 & 242 \\ \hline
      North &  IR & F182M - F430M & 0.60 & 121\\
      North & Pa$\alpha$ & F182M - F187N & 0.10 & 164 \\
      North & Br$\alpha$ & F430M - F405N & 0.10 & 37 \\ \hline
      South &  IR & F200W - F444W & 0.60 & 456 \\
      South & Pa$\alpha$ & F200W - F187N & 0.10 & 161 \\
      South & Br$\alpha$ & F444W - F405N & 0.10 & 71 \\ 
      South & H$_2$ & F444W - F470N & 0.10 & 70 \\ 
      South & H$_{2\,(2.12 \mu \text{m})}$  & F200W - F212N & 0.10 & 193 \\ \hline 
    \end{tabular}
    \label{tab:excess_selections}
\end{table}

\subsection{Colour-colour selections}

\begin{figure}
    \centering
    \includegraphics[width=0.49\textwidth]{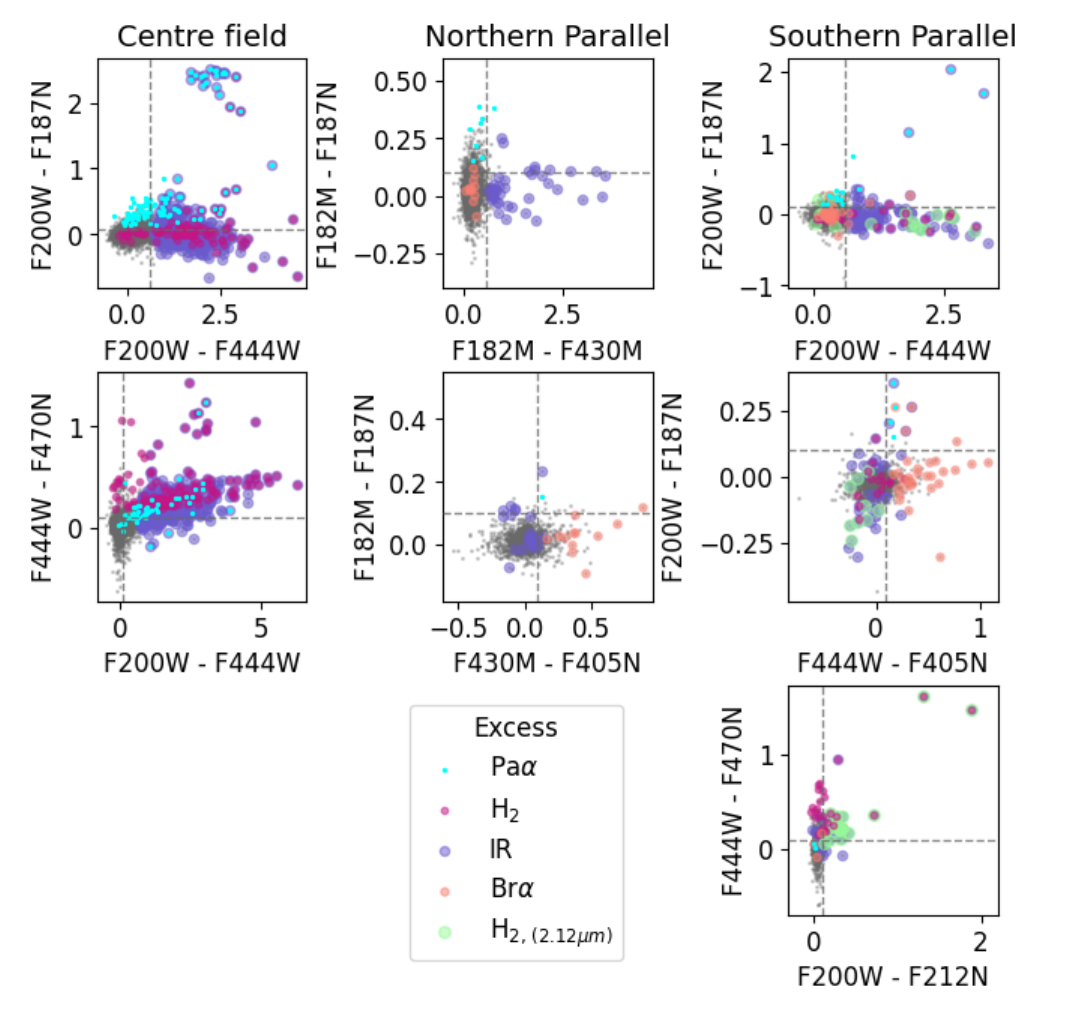}
    \caption{Colour-colour diagrams used to select sources with excess emission due to warm dust or emission lines. \textit{Left panels}: Selection of sources in the centre field with Pa$\alpha$ (cyan), H$_2$ (magenta), and IR (purple) excess. \textit{Middle panels}: Selection of sources in the northern parallel field with Pa$\alpha$ (cyan), Br$\alpha$ (peach), and IR (purple) excess. \textit{Right panels}: Selection of sources in the southern parallel field with Pa$\alpha$ (cyan), H$_2$ (4.70 $\mu$m, magenta), H$_{2}$ (2.12 $\mu$m, pale green), Br$\alpha$ (peach), and IR (purple) excess.}
    \label{fig:CCD_selections}
\end{figure}

While we made the initial selections based on only one colour, the position of excess sources in colour-colour space gives further insights into their nature. Figure \ref{fig:CCD_selections} shows different colour combinations for the different fields. We note that combining different colours to create these figures significantly reduces the number of sources, especially when one of the long-wavelength channel narrow-band filters is included because these data are not as deep.

The leftmost panels of this figure show the colour-colour combinations of the centre field, revealing a population of stars that have both excess in the IR and Pa$\alpha$ emission. These are interesting candidates for accreting still deeply embedded PMS stars. The right panel in Fig. \ref{fig:CMD_center_excess} shows that many of them are indeed located where PMS stars are expected, but some brighter stars also have excess in IR and Pa$\alpha$. These could be candidates of more massive PMS stars. We note again the population of stars with very high Pa$\alpha$ excess (F200W - F187N > 2), which also have a significant IR excess. As described above, many of them show extended structures in the F187N filters and are discernible as mini-HII regions in the images.

Similarly, there are stars with both an excess in IR and H$_2$ emission in the centre field. Some of them are also Pa$\alpha$-excess sources. De Marchi et al. (2023) showed with NIRSpec spectroscopy that accreting PMS stars might also show H$_2$ lines from their natal cloud. Inspection of the sources in the F090W - F200W CMD (Fig. \ref{fig:CMD_center_excess}) finds them to be brighter than our isochrone-based PMS stars, likely because of the limited depth of the long wavelength data. These stars are likely more massive YSOs and appear to fall on the 1 - 2 Myr isochrones shown in Fig. \ref{fig:isochrones} for example. 
Interestingly, all sources with excess in both Pa$\alpha$ and H$_2$ also show IR excess.

Because there are fewer stars and the distance to the centre of 30\,Doradus is larger, the colour-colour diagrams for the northern and southern parallel observations are less strongly populated with excess sources. The southern parallel observations include a handful of sources with both Pa$\alpha$ and IR excess emission, as well as some with H$_2$ emission, creating an excess in both the F470N and F212N filters. We find that sources with an excess in Pa$\alpha$ do not necessarily also show an excess in Br$\alpha$. This probably is because Br$\alpha$ is intrinsically fainter than Pa$\alpha$ and also because the F444W filter contains the line, hindering the characterisation of the continuum. 

\section{Spatial distributions of different populations}
\label{sect:spatial_dist}
In the following, we focus our analysis on the spatial distribution of the different populations across the field of view. We then also compare to the distribution of cold molecular gas.


\subsection{Two-dimensional histograms of spatial distributions}

\begin{figure*}
    \centering
    \includegraphics[width=0.98\textwidth]{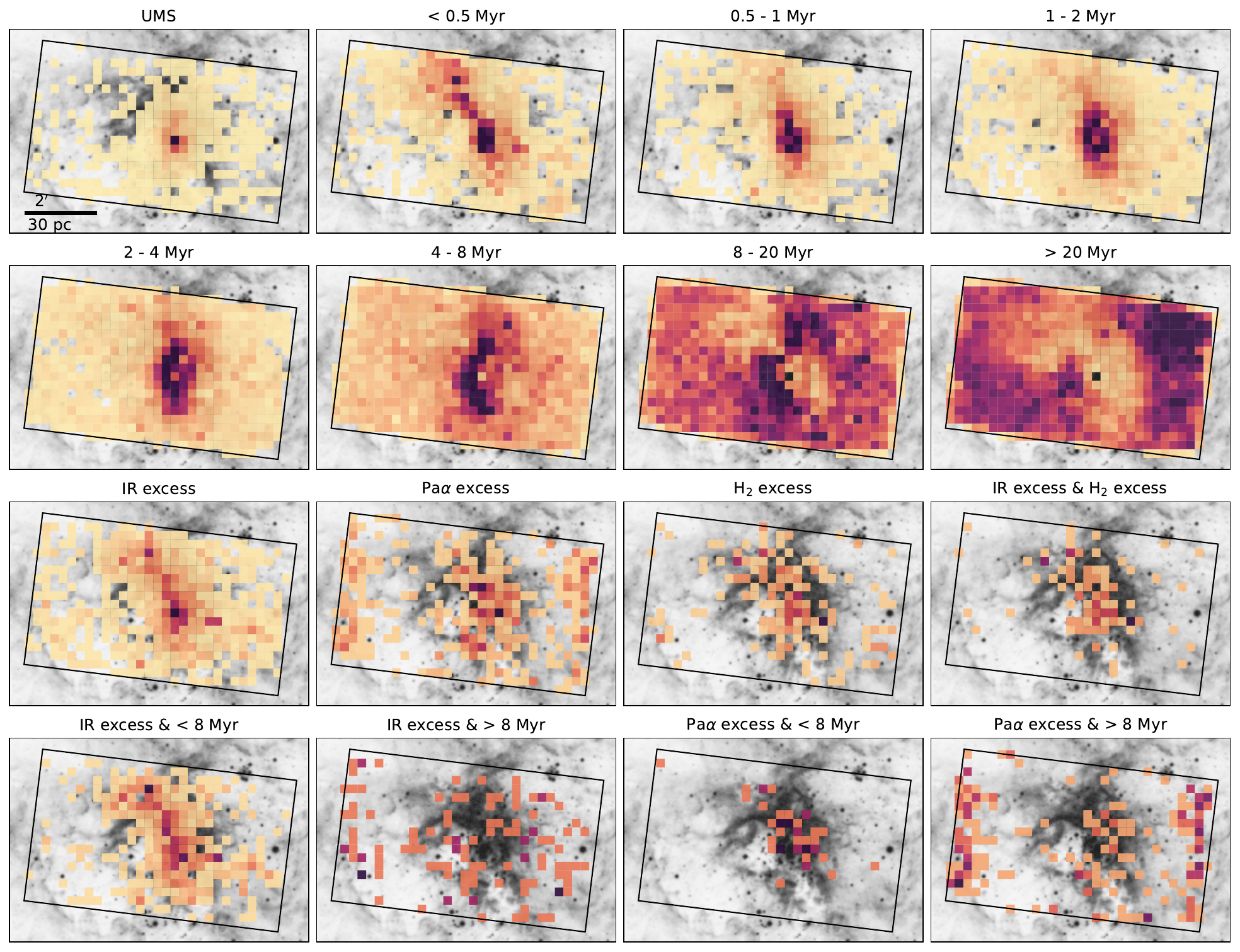}
    \caption{Spatial distributions of different populations of the centre field as two-dimensional histograms with 15\arcsec\,bin size overlaid on the DSS2 red greyscale image. \textit{Top and first row}: Selections based on the F090W - F200W colour magnitude diagram and comparisons with isochrones as shown by two-dimensional histograms. \textit{Third row}: Histograms show individual stars selected based on their excess. \textit{Bottom row}: Histograms showing stars that fall in two or more selections. A scale bar of 2\arcmin\,is shown in the first panel.}
    \label{fig:spatial_dist}
\end{figure*}

\begin{figure}
    \centering
    \includegraphics[width=0.49\textwidth]{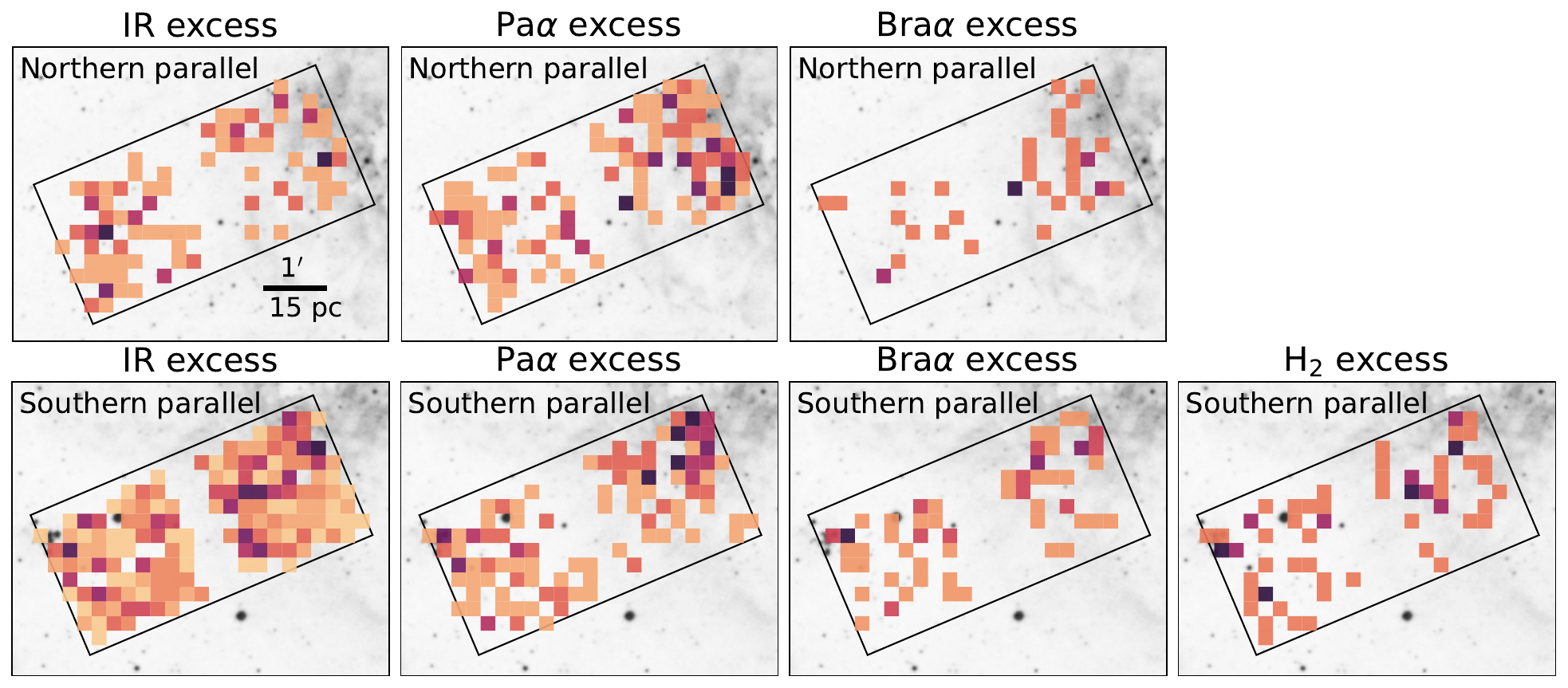}
    \caption{Spatial distributions of different populations of the northern (top) and southern parallels (bottom) as two-dimensional histograms with 15\arcsec\,bin size overlaid on the DSS2 red greyscale image (see Fig. \ref{fig:footprint} for the placement of these fields in the 30\,Doradus region). Histograms show individual stars selected based on their excess.}
    \label{fig:spatial_dist_parallels}
\end{figure}

Figure \ref{fig:spatial_dist} shows the spatial distributions of the various selected populations in the centre field as two-dimensional histograms overplotted on the DSS2 red image for visualisation. Each bin in the histograms corresponds to 15\arcsec $\times$ 15\arcsec. Figure \ref{fig:spatial_dist_parallels} shows similar histograms for the parallel fields.

The distribution of UMS stars peaks sharply in the centre of R\,136, the young massive cluster in the centre of the 30\,Doradus region, and shows a rather smooth distribution across the field of view. This underlines the findings by the HTTP and the VFTS that young, massive stars are widely distributed (e.g. \citealt{Sabbi2013, Schneider2018}), likely indicating a widely distributed star formation within the last million years. The very central bin on top of the centre of R\,136 shows a drop in these PMS stars, however, we find that this because the extremely massive stars in the centre of R\,136 are saturated in all of the NIRCam filter and are thus lacking from our catalogue.

In contrast, the youngest and reddest selected PMS stars (\mbox{$<$ 0.5 Myr}) show an elongated distribution towards the north-east of R\,136, reaching deep within the highly extinguished region. Ground- and space-based NIR detections of PMS stars have already established that star formation is ongoing in this region (e.g. \citealt{Rubio1998, Walborn1999}). With increasing age of the isochrones used for selection, this elongated structure becomes weaker. Instead, stars with colours consistent with ages between 1 and 4 Myr are again concentrated on R\,136, but not as peaked as the UMS stars. Stars older than 4 Myr begin to show an elongated curved structure that is displaced from the centre of R\,136 and instead fills the rather dust-free region in the centre of NGC\,2070. This structure is also found in the 8 - 20 Myr age bin, but disappears for the oldest stars ($>$ 20 Myr), where the dusty regions instead become visible as regions of lower density. We found this to be caused by a selection effect because this population mainly contains faint stars that have a lower detection rate in these regions due to extinction. An additional magnitude threshold of F090W < 22.5 mag for these stars results in a smooth distribution, indicating that this population belongs to the LMC field population.

\begin{figure*}
    \centering
    \includegraphics[width=0.98\textwidth]{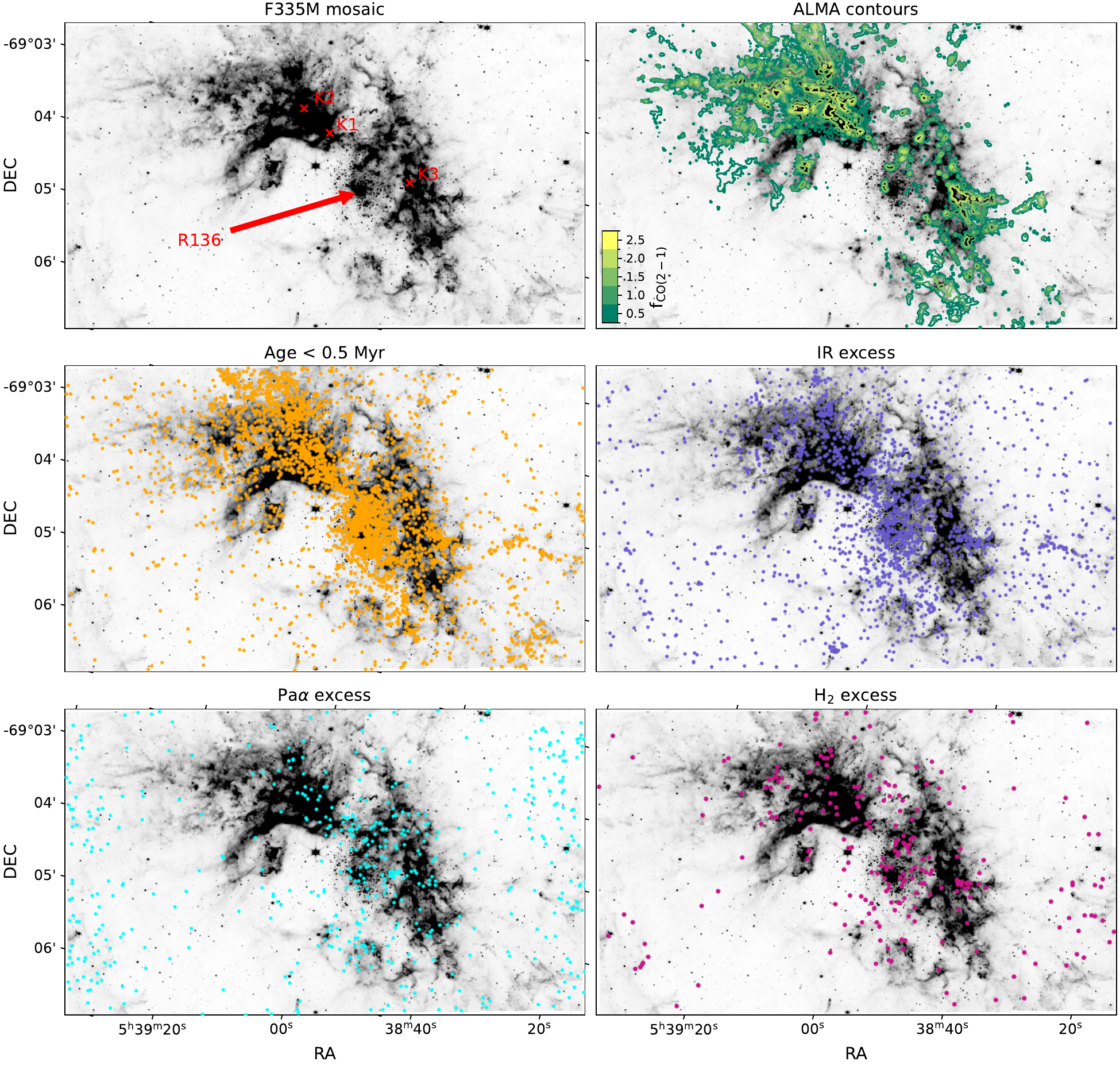}
    \caption{Spatial distribution of different populations in the centre field on top of the F335M mosaic (\textit{top left} and as background in each panel). The red arrow indicates the location of the central cluster R\,136, and the red crosses mark the positions of previously well-studied IR-bright knots (e.g. \citealt{Walborn1999}). \textit{Top right}: $^{12}$CO moment zero maps from ALMA (data from \citealt{Wong2022}) as green to yellow contours.  The coloured dots show individual stars selected in different populations. The image has a field of view of 7.4\arcmin\,$\times$ 4.4\arcmin\,(110 $\times$ 66 pc).}
    \label{fig:spatial_dist_h2}
\end{figure*}

The third row of Fig. \ref{fig:spatial_dist} shows two-dimensional histograms for sources selected based on their excess emission. The IR-excess sources show a very similar elongated distribution as the youngest PMS stars ($<$ 0.5 Myr). 
The Pa$\alpha$-selected stars reach the highest densities in the central region of 30\,Doradus and overlap with the regions that also have a high density of young and IR-excess populations. Additionally, Pa$\alpha$-excess sources appear at the edges of the nebular regions at the borders of the central NIRCam field of view. Sources with H$_2$ excess are concentrated towards the centre of 30\,Doradus and appear to trace the molecular filaments (see also Fig. \ref{fig:spatial_dist_h2}). The parallel fields also contain these excess sources, even as far as 100 kpc from R\,136 (Fig. \ref{fig:spatial_dist_parallels}).

In the bottom row of Fig. \ref{fig:spatial_dist}, we show histograms when two or more selections are combined. We find that sources that both have an IR excess and are selected to be redder than the 8 Myr isochrone in the F090W - F200W CMD are clearly elongated towards the north-east, while IR-excess stars with older ages are more uniformly distributed across the field.
This highlights that the youngest PMS stars often also have IR excess, indicating that they have very red spectral energy distributions. This is expected for PMS stars that are still deeply embedded in their natal cloud (e.g. \citealt{Robitaille2007}).

Sources with Pa$\alpha$ excess and ages younger than 8 Myr are found to be located on regions with high nebulosity in the central region, while almost all Pa$\alpha$ excess stars with older ages are located in the other regions of the cloud. We suggest that this indicates that the radiative feedback from the massive stars in R\,136 is efficient in destroying the accretion disks of PMS stars in the central regions, while PMS stars at larger radii are able to accrete longer. This erosion of accretion disks due to photo-evaporation was also suggested by \cite{DeMarchi2011} based on the finding that older accreting PMS stars are displaced from the centre of R\,136 based on H$\alpha$ narrow-band HST observations of a central field in 30\,Doradus.

\subsection{Comparison with cold molecular gas}
\label{sect:molecular_gas}

While Fig. \ref{fig:spatial_dist} compares the spatial distributions of different selected populations against the appearance of 30\,Doradus in the optical, we compared the youngest PMS stars and the excess sources in the centre field to the ALMA $^{12}$CO(2$-$1) moment zero map from \cite{Wong2022}. As discussed in detail by these authors, the high-resolution CO maps show a high degree of complexity, with fine filaments organised in a larger bow-tie structure that fans outwards from the centre of R\,136. By comparing with the NIRCam images, we find that the F335M mosaic matches the CO maps best, likely because this filter traces the emission of polycyclic aromatic hydrocarbon molecules (PAHs) that closely follow the distribution of the molecular gas. 

The comparison to the youngest PMS shows that many of them are located in cold molecular gas filaments, especially towards the north.
The overlap for the IR-excess sources is slightly less obvious because many sources outside the CO filaments are also found. Additionally, both populations show clumps and substructure, many of which are associated with features in the CO map. In addition, there appears to be a diffusely distributed component to these populations, especially the IR-excess sources, illustrating that recent star formation in 30\,Doradus was not only restricted to the densest regions. 

Similarly, many stars with Pa$\alpha$ excess are found outside and at the edges of the cold molecular regions, but some are also located in the nodes of high (but not highest) flux density. 
In contrast, some of the H$_{2}$-excess sources are found in the highest flux density regions, likely indicating embedded regions. Many of the H$_{2}$-excess sources are also found outside of the cold molecular gas filaments, however. 

\begin{figure*}
    \centering
    \includegraphics[width=0.95\textwidth]{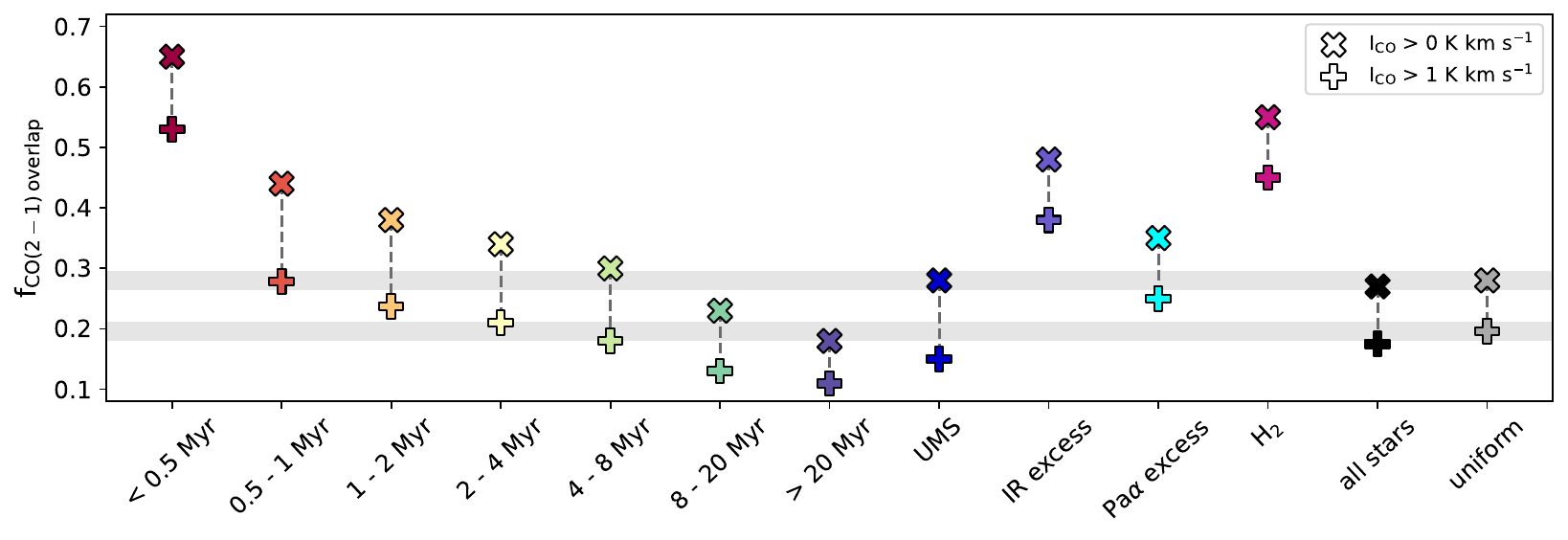}
    \caption{Fraction of sources in different selections that overlap with the ALMA CO(2$-$1) map from \cite{Wong2022}. The cross shows the fraction of overlap without any intensity constraint, the plusses show the overlap with regions in which the CO intensity exceeds 1.0 K km s$^{-1}$. The black symbols refer to the full JWST catalogue, and the grey symbols and bands show the expected values for a uniform distribution of stars across the field of view with both JWST and ALMA coverage. These values are higher than for the old stars (> 20 Myr) because only a few are detected in the highly extinguished regions.}
    \label{fig:CO_selection}
\end{figure*}

To determine the overlap between populations and the CO maps, we determined the number of stars that are located in a non-zero entry of the ALMA CO(2$-$1) map compared to all stars in the footprint of the CO map as seen in Fig. \ref{fig:CO_selection}. We also show the fraction of stars that overlap with regions in which the CO flux densities $\geq$ 1 K km s$^{-1}$ as a stricter limit.

The highest fractions are reached for stars that are selected to be younger than 0.5 Myr. Here, 65 \% of these stars are associated with CO, and 53 \% even fall in the high flux density regions. This underlines the qualitative finding based on their distribution on the CO maps and suggests that these PMS stars are indeed still associated with the cold molecular gas filaments.

The next-highest fraction is found for the stars selected based on the H$_2$ excess as identified by the F470N narrow-band filter. Even though this filter traces emission from excited molecular hydrogen, it appears to still be a viable tracer of the most embedded regions. Similarly, selecting sources with IR excess from broad-band colours results in a high overlap with the CO maps.

For the other selections, we find that the fraction with CO overlap decreases with age to fractions $< 20$ \% for the oldest stars. While this small fraction shows that the old population is unlikely to be associated with the CO emission, we caution that this small fraction is also driven by the low luminosity of these stars. As the CO emission overlaps with the regions of highest extinctions, we detect fewer of these faint, old stars in these regions, as also shown in Fig. \ref{fig:spatial_dist}. To illustrate this further, we include in Fig. \ref{fig:CO_selection} the values for all stars in the JWST catalogue. Of these, 27 \% fall onto a non-zero entry of the CO map, and 18 \% lie in a region with $\geq$ 1 K km s$^{-1}$. When we instead assume a completely uniform distribution of stars, 28 \% would be associated with non-zero CO and 20 \% with the higher-intensity regions. With a magnitude limit on the old stars (> 20 Myr), we can recover these fractions. This indeed suggests a uniform distribution.

Although this simple comparison cannot account for the three-dimensional distribution in 30\,Doradus, it already shows that the youngest populations are not all associated with the dense molecular gas as traced by CO. This could either be an observational effect due to the limited depth of the CO observations, or it might indicate a large fraction of CO-dark H$_2$, as also concluded by \cite{Nayak2023}. These authors compared the location of massive YSO candidates to the same CO(2$-$1) map and reported that only 38\% were associated with the CO molecular gas. 

In this regard, the sources with H$_2$ excess that are not associated with the CO molecular gas are particularly interesting because the excited H$_2$ gas traced by the 4.70 $\mu$m line might be directly associated with an accretion disk around the stars and not with the natal molecular cloud, as confirmed with NIRSpec spectroscopy of some PMS stars in NGC\,346 (De Marchi et al.2023). However, spectroscopy of multiple H$_2$ lines is required to determine the excitation mechanism (e.g. \citealt{Jones2022}).

\section{Discussion}
\label{sect:discussion}
We have presented an photometric analysis of JWST NIRCam data in three different fields in the 30\,Doradus star-forming region. Using public data of the central field as well as two NIRCam pointings conducted as NIRSpec parallel observations, we derived PSF photometry for more than 200000 stars detected in at least two NIRCam filters. We used selected different populations by a) comparing with theoretical isochrones and b) applying colour-cuts to select sources with a colour excess due to emission from warm dust or excited emission lines. We then analysed the spatial distributions of different populations. 

\subsection{Comparing approaches based on optical HST and JWST data}
The high sensitivity and spatial resolution of NIRCam allowed us to analyse CMDs with similar techniques and accuracy as is typically used for HST photometry. We found a good match to expectations from theoretical isochrones after dereddening the data of the centre field, and we used these isochrones to select PMS stars in a F090W - F200W CMD with ages between 0.5 and 20 Myr. We note here that we did not fit for the ages, but rather used their location in the reddening-corrected CMD to select stars that would fall in a given age bin. Our selection reaches F090W $\sim$ 25 mag, selecting stars with masses between $\sim 0.1$ and 3 $M_\sun$. This mass range is similar to that studied with HST within the HTTP project \cite{Sabbi2013, Sabbi2016, Ksoll2018}. Similar to the results of \cite{DeMarchi2011}, \cite{Sabbi2016}, and \cite{Ksoll2018}, we found that the young PMS stars are centrally concentrated and sometimes organised in loose clumps and filaments, while older stars are displaced from the core of R\,136. The distribution of UMS stars is strongly peaked at the location of R\,136 but also shows a widely distributed component. This agrees with the distribution of UMS stars in the HTTP \cite{Sabbi2013} and the finding of widely distributed O and B stars from the VFTS \citep{Schneider2018}. 

As an alternative to using broad-band colours alone, \cite{DeMarchi2010} established that low-mass PMS stars can be identified with high confidence by using optical HST colours in combination with H$\alpha$ narrow-band observations to search for H$\alpha$ excess emission. Created by the gravitational energy released from infalling matter in an accretion disk around PMS stars, this feature can be clearly associated with low-mass star formation and has now long been used to identify PMS stars in various star-forming regions in the Local Group (e.g. \citealt{DeMarchi2010, DeMarchi2013, Tsilia2023, Vlasblom2023}). Using HST WFC3 data, \cite{DeMarchi2011a} and \cite{DeMarchi2017} used this approach to study the PMS stars with H$\alpha$ excess in 30\,Doradus, finding them to be centrally concentrated. With the JWST, we could now perform a similar approach by using the excess in Pa$\alpha$ as traced by the F187N filter. As also found by \cite{DeMarchi2017}, the Pa$\alpha$ sources appear to have a broad range in ages. Some of them are clearly located in the PMS region of the CMD, while others are close to the main sequence. This agrees with the interpretation that PMS stars older than 20 Myr can still be accretion material. Based on NIRSpec spectroscopy, emission lines including Pa$\alpha$ were indeed recently identified in PMS stars in NGC\,346 in the SMC (De Marchi et al. 2023), and a similar study is planned for a sample of about 100 PMS stars in 30\,Dor. 

Using filters also at longer wavelengths, we were further able to select stars from the same data with an excess in infrared colours, thereby tracing stars with excess emission due to warm dust. This allowed us to study these low-mass PMS stars even in the highly dust obscured regions of 30\,Doradus. This is better than what was previously possible in the optical with HST. Comparing the distributions of CMD-selected PMS stars and IR-excess sources, we found the latter to be distributed farther out in the optically obscured regions of the nebula. 
Using NIRCam observations of the SMC star-forming region NGC\,346, \cite{Jones2023} also found that the JWST-selected PMS stars closely follow the dust filaments. Compared to this region, the PMS stars in 30\,Dor appear to be both more widely distributed and numerous. This might be expected because 30\,Doradus is a more massive region and hosts the most energetic starburst event in the Local Group. 

\subsection{Comparison to NIR studies}
Over the past decades, 30\,Doradus has been studied extensively in the near-infrared regime using both ground-based telescopes and HST NICMOS. Numerous IR-bright sources have been detected, in particular towards the north-east and in the western filament next to R\,136 (e.g. \citealt{Hyland1992, Rubio1992, Rubio1998, Walborn1999, Brandner2001}). In these regions, we now also identify a large number of both IR excess and young PMS stars. A particular focus was placed in the previous studies on some of the brightest sources: three early O-type stars embedded in nebular knots, called knots 1, 2, and 3 \citep{Walborn1987, Walborn1991}. HST NICMOS observations of these regions revealed them to consist of multiple stellar components embedded in complex nebular and dusty structures, and they are surrounded by fainter stars \citep{Walborn1999, Brandner2001}. Follow-up observations with HST WFPC2 further showed the complex nebular structure and embedded populations around these knots \citep{Walborn2002}.

\begin{figure}
    \centering
    \includegraphics[width=0.48\textwidth]{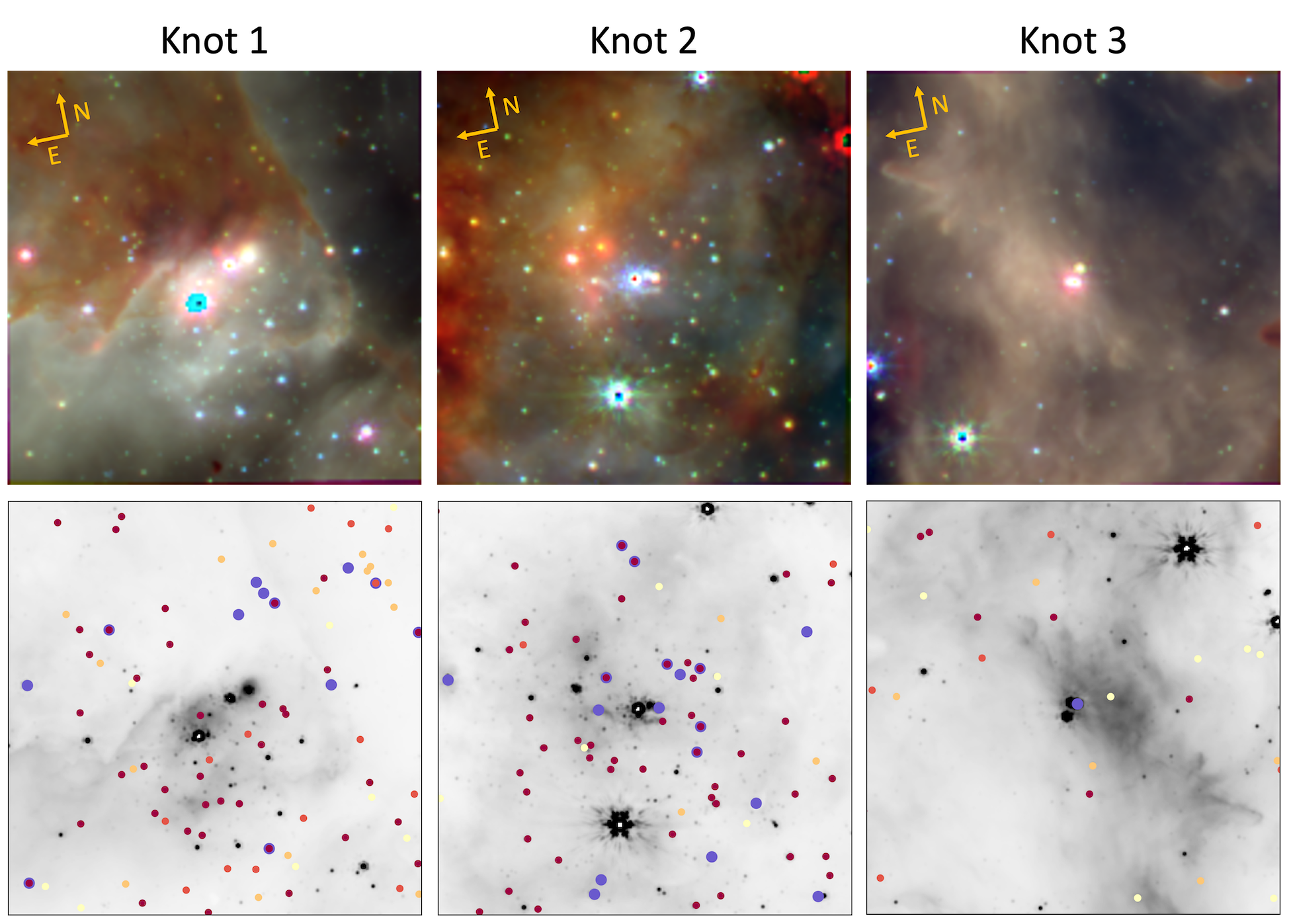}
    \caption{Zoom-ins around the three IR-bright knots 1, 2, and 3. \textit{Top row:} Colour images of 5\arcsec\,$\times$ 5\arcsec regions. Red: F444W, green: F200W, blue: F090W. \textit{Bottom row:} Greyscale F200W images with selected sources as coloured dots. The larger purple dots refer to IR-excess sources, and the smaller maroon, red, orange, and yellow dots are stars selected with isochrones with ages of 0.5, 1, 2, and 4 Myr, respectively. }
    \label{fig:knots}
\end{figure}

Figure \ref{fig:knots} shows the JWST view of these three regions, presented in small cut-outs around each knot (see Fig. \ref{fig:spatial_dist_h2} for their position in the field of view). The panels in the bottom row show the position of sources selected with isochrones (up to 4 Myr) and with IR excess. While the originally studied O-stars are saturated in at least one of the JWST filters, many of the fainter bluer sources seen close to them are young PMS stars. The sources that appear orange or red in the RGB images are predominantly IR-excess sources in our selection. Additionally, the long-wavelength filter F444W shown in red in these images reveals several sources that are not visible at shorter wavelengths.

\subsection{Comparison to Spitzer and Herschel sources}

Previous studies in the infrared with \textit{Spitzer} or \textit{Herschel} were limited to bright, massive YSOs, but used similar colour-based approaches as in this work, or a spectral energy distribution fitting to select PMS stars in the embedded stages of star formation (e.g. \citealt{Whitney2008, GruendlChu2009, Carlson2012}). To understand how these studies compare to our results, we matched our photometric catalogues to the 299 YSO candidates from \cite{Nayak2023}. This matching is not straightforward because our catalogues contain many sources within the photometric uncertainties of these candidates. Therefore, we first selected all stars within 1\arcsec\, from the YSO candidates and then performed a match when a source with an F335M magnitude within 1 mag of the IRAC 3.6 $\mu$m or with an F444W magnitude within 1 mag of the IRAC 4.5 $\mu$m magnitude was reported in \cite{Nayak2023}. This resulted in 72 matches out of the 488 sources listed in \cite{Nayak2023}, 30 of which were classified as YSOs. 

We note here that not all sources of \cite{Nayak2023} overlap the NIRCam footprint, but the main reason for the low number of matches probably is that most of the YSO candidates are so bright that they are saturated in the NIRCam imaging. The matched sources exclusively have 2MASS K$_s$ magnitudes $> 15$, while the brighter sources are not recovered in the NIRCam images. Consequently, in our selections, even the matched YSOs are mostly missed due to their brightness, showing that an analysis with JWST like this is highly complementary to previous studies of higher mass YSOs. Additionally, in some cases, the YSO candidates could not be matched in magnitude even when measurements were available. The reason probably is that in some cases, the JWST imaging reveals that the YSO candidates consist of more than one source (see also e.g. \citealt{Jones2022}). 

\subsection{30\,Doradus in the context of star cluster formation}

As some of the oldest stellar systems in the Universe, ancient globular clusters (GCs) are often used as tracers of galaxy evolution. These tracers allow us to constrain the assembly of galaxies (e.g. \citealt{Brodie2006, Beasley2020}). However, because they are so old, it is unclear so far how they formed. In general, two pathways are discussed, which either propose that GCs formed under conditions unique to the high-redshift Universe, for example in dark matter minihaloes, or that they are simply the product of violent star formation in high-density conditions that are rare in the local Universe (e.g. \citealt{Forbes2018}). In the latter scenario, young massive clusters (YMCs) have been proposed as local analogues of GCs at high redshifts and seem to have similar properties as GCs in terms of masses and densities (e.g. \citealt{Portegies2010, AdamoBastian2018, Adamo2020_review}). However, their formation is rare in the Local Group, and they are found in high numbers in interacting galaxies (e.g. \citealt{Whitmore2010, Adamo2020}).

In this context, NGC\,2070 in 30\,Doradus is a unique region in which to study the formation of massive star clusters in great detail. While there are other YMCs in the Milky Way or the Small Magellanic Cloud (e.g. Westerlund 1 \& 2), NGC\,2070 with its subcluster R\,136 is by far the most massive YMC. Mass estimates of the full cluster range from $\sim$ 7 $\times 10^4$ M$_\sun$ to 5 $\times$ 10$^{5}$ M$_\sun$, while the R\,136 subcluster has an estimated mass between $2.2 \times 10^4$ and 1 $\times$ 10$^{5}$ M$_\sun$ \citep{Hunter1995, Andersen2009, Bosch2009, Cignoni2015}. 

\begin{figure}
    \centering
    \includegraphics[width=0.45\textwidth]{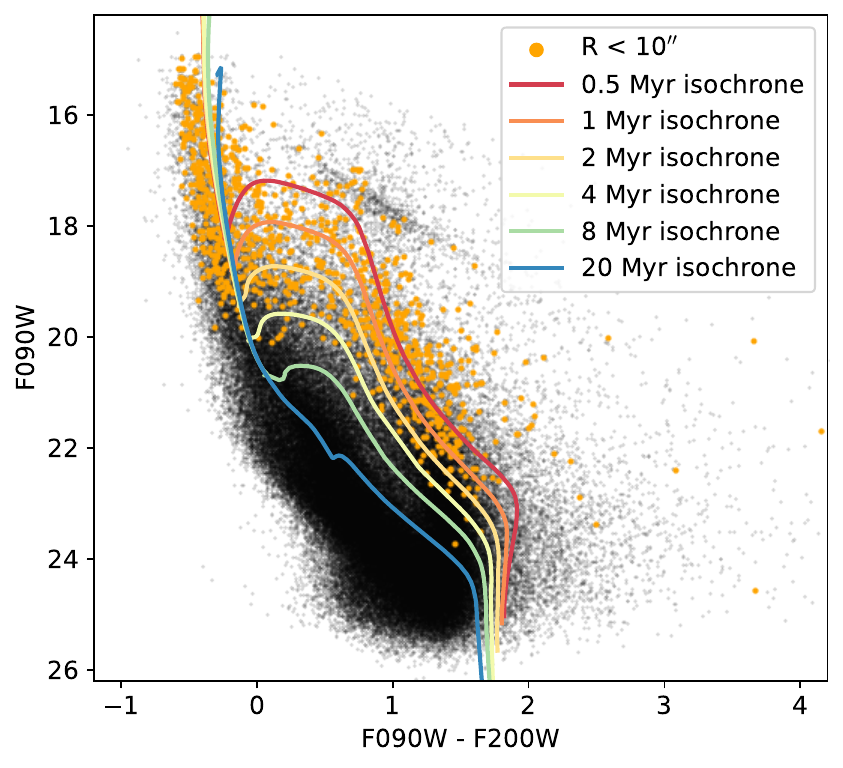}
    \caption{Colour-magnitude diagram of R\,136 (orange dots) on top of the full region (black). The orange dots refer to all stars within 10\arcsec\,from the centre of R\,136. The coloured lines show theoretical MIST isochrones.}
    \label{fig:CMD_R136}
\end{figure}

R\,136 is known to have an age of of 0.5 - 2 Myr based on its CMD, and the older more diffuse population in NGC\,2070 has an age of $\sim$ 3 - 7 Myr \citep{Selman1999, Sabbi2012, Cignoni2015, Bestenlehner2020, Brands2022}. Fig. \ref{fig:CMD_R136} shows that where we selected stars within a 10\arcsec\,distance (2.5 pc, the typical effective radius of a GC) from the centre of R\,136, we can reproduce this finding with the JWST photometry. \cite{Dominguez2023} were able to explain the presence of the younger R\,136 subcluster within NGC\,2070 with N-body simulations, finding that the original feedback from the stars that formed in the first generation was likely not sufficient to fully expel the gas, allowing a second generation to form that was more centrally concentrated. They interpreted this within the frequently discussed multiple population problem of GCs (e.g. \citealt{Bastian2018}). 

The elongated structure of very young ($<$ 0.5 Myr), IR-bright stars towards the north-east of the cluster might present additional evidence for a still ongoing hierarchical formation of this massive star cluster. As hydrodynamical simulations suggest, small subclusters form first in filaments during the collapse of giant molecular clouds and then grow through hierarchical mergers while still accreting gas to form a more massive star cluster on a timescale of a few million years (e.g. \citealt{Howard2018, Lahen2020, Guszejnov2022}). In this context, the observed CO filaments might correspond to inflowing material that supplies cold molecular gas and the PMS stars embedded within to the central region of the star cluster to sustain star formation. 
At some point, feedback from massive stars is thought to heat and dispel the molecular gas, slowing down star formation (e.g. \citealt{Krumholz2019, Krause2020}). Given the radiation pressure from the young massive stars in R\,136, the filamentary CO structure could also be interpreted as stemming from outflows \citep{Wong2022}. Moreover, our finding that the older Pa$\alpha$ sources are mainly found at larger distances from R\,136 might indicate that the feedback from the massive stars in R\,136 already impedes the accretion of PMS stars in the centre. 
When the first supernovae explode in NGC\,2070, the expelled gas and star formation will likely be halted, leaving a mostly gas-free cluster with stars that formed within a very short time span of a few million years.

\section{Conclusions}
\label{sect:conclusions}
We presented a JWST NIRCam analysis of the young and old populations in the 30\,Doradus star-forming region. We focused on how JWST filters can be used to select different populations, and we analysed their spatial distributions in the region. Our main results are summarised below.

\begin{itemize}
    \item We presented PSF photometry for more than 200000 sources with at least two NIRCam magnitudes, about 150000 in the centre field and the remaining sources in two parallel fields.
    Including a short-wavelength filter bluer than 2 $\mu$m is crucial for separating different populations in CMDs using classical approaches based on isochrones. We found that the same methods that are long established with HST photometry can be applied. The PSF photometry closely matches the expectations from theoretical isochrones after a reddening correction is applied.
    
    \item We used different approaches to separate various populations, such as a comparison with isochrones of different ages and colour-cuts to separate populations that show excess emission due to either warm dust as traced by a long-wavelength broad-band filter or due to line emission from excited neutral and molecular hydrogen as traced by a corresponding narrow-band filter. 
    
    \item Using isochrones on top of the F090W - F200W CMD, we selected low-mass sources whose colours are consistent with ages between 0.5 and 20 Myr down to masses $\sim 0.1\,M_\sun$. We found that the youngest sources ($< 0.5$ Myr) show an elongated structure towards the north-east of the region that extends deep into the nebular regions. Stars with ages between 1 and 4 Myr are centrally concentrated on R\,136. In contrast, PMS stars with intermediate ages between 4 Myr and 20 Myr were found to be spatially separated from the very centre of R\,136, in an elongated structure, and stars older than 20 Myr appear to be uniformly distributed across the 30\,Doradus region, indicating that they belong to the LMC field population. 

    \item Selecting stars that show an excess towards long wavelengths using the F200W - F444W colour, we found a very similar distribution as for the isochrone-selected $< 0.5$ Myr PMS stars in a more elongated structure. This suggests that there is a population of still embedded very young PMS stars to the north-east of the central cluster that possibly still feeds its assembly through hierarchical mergers of small subclusters.
    
    \item Using the narrow-band filter F187N, we selected sources with an excess in Pa$\alpha$ as a tracer of ongoing accretion onto the PMS star. These sources are concentrated towards the centre of the cluster, but are also found further outwards in the optically visible nebular features of 30\,Doradus. Placing the Pa$\alpha$ sources in a colour-magnitude diagram showed that they are not only in the PMS locations, but also closer to the main sequence, revealing that the accretion can persist many tens of million years. The F470N narrow-band filter as an indicator for excess due to H$_2$ emission predominantly selects more massive stars because the depth is limited. These sources appear to be distributed along the dust regions, but are also found to be distributed more widely.

    \item While most of the selected PMS and YSOs are found in the central field, we also find sources with excess due to warm dust and emission lines in the parallel fields. This suggests that star formation is ongoing or very recent as far as 100 pc from the centre. The nebulosity seen in the colour images supports this.
    
    \item By comparing with available CO(2$-$1) maps from ALMA \citep{Wong2022}, we found that most of the youngest PMS stars (< 0.5 Myr) are associated with the CO molecular gas, whereas older or IR-excess stars do not always overlap with the CO maps. About half of the H$_2$-excess sources do not appear to be associated with CO emission. We interpreted this as further evidence for a large amount of CO-dark H$_2$ gas.
\end{itemize}
JWST NIRCam, with its many filter choices and enhanced sensibility, has allowed us to combine methods that were previously used separately to perform a comprehensive study of the star formation in 30\,Doradus.

\begin{acknowledgements}
      We thank an anonymous referee for comments and suggestions that have helped to polish this paper.
      KF acknowledges support through the ESA research fellowship programme. Based on observations with the NASA/ESA {\em Hubble Space Telescope} and the NASA/ESA/CSA {\em James Webb Space Telescope}, which are operated by AURA, Inc., under NASA contracts NAS5-26555 and NAS 5-03127. This work made use of Astropy:\footnote{\url{http://www.astropy.org}} a community-developed core Python package and an ecosystem of tools and resources for astronomy \citep{astropy2013, astropy2018, astropy2022}
\end{acknowledgements}

\bibliographystyle{aa} 
\bibliography{References}

\begin{thebibliography}{87}
\expandafter\ifx\csname natexlab\endcsname\relax\def\natexlab#1{#1}\fi

\bibitem[{{Adamo} \& {Bastian}(2018)}]{AdamoBastian2018}
{Adamo}, A. \& {Bastian}, N. 2018, in Astrophysics and Space Science Library,
  Vol. 424, The Birth of Star Clusters, ed. S.~{Stahler}, 91

\bibitem[{{Adamo} {et~al.}(2020{\natexlab{a}}){Adamo}, {Hollyhead}, {Messa},
  {Ryon}, {Bajaj}, {Runnholm}, {Aalto}, {Calzetti}, {Gallagher}, {Hayes},
  {Kruijssen}, {K{\"o}nig}, {Larsen}, {Melinder}, {Sabbi}, {Smith}, \&
  {{\"O}stlin}}]{Adamo2020}
{Adamo}, A., {Hollyhead}, K., {Messa}, M., {et~al.} 2020{\natexlab{a}}, \mnras,
  499, 3267

\bibitem[{{Adamo} {et~al.}(2020{\natexlab{b}}){Adamo}, {Zeidler}, {Kruijssen},
  {Chevance}, {Gieles}, {Calzetti}, {Charbonnel}, {Zinnecker}, \&
  {Krause}}]{Adamo2020_review}
{Adamo}, A., {Zeidler}, P., {Kruijssen}, J.~M.~D., {et~al.} 2020{\natexlab{b}},
  \ssr, 216, 69

\bibitem[{{Andersen} {et~al.}(2009){Andersen}, {Zinnecker}, {Moneti},
  {McCaughrean}, {Brandl}, {Brandner}, {Meylan}, \& {Hunter}}]{Andersen2009}
{Andersen}, M., {Zinnecker}, H., {Moneti}, A., {et~al.} 2009, \apj, 707, 1347

\bibitem[{{Astropy Collaboration} {et~al.}(2022){Astropy Collaboration},
  {Price-Whelan}, {Lim}, {Earl}, {Starkman}, {Bradley}, {Shupe}, {Patil},
  {Corrales}, {Brasseur}, {N{\"o}the}, {Donath}, {Tollerud}, {Morris},
  {Ginsburg}, {Vaher}, {Weaver}, {Tocknell}, {Jamieson}, {van Kerkwijk},
  {Robitaille}, {Merry}, {Bachetti}, {G{\"u}nther}, {Aldcroft},
  {Alvarado-Montes}, {Archibald}, {B{\'o}di}, {Bapat}, {Barentsen},
  {Baz{\'a}n}, {Biswas}, {Boquien}, {Burke}, {Cara}, {Cara}, {Conroy},
  {Conseil}, {Craig}, {Cross}, {Cruz}, {D'Eugenio}, {Dencheva}, {Devillepoix},
  {Dietrich}, {Eigenbrot}, {Erben}, {Ferreira}, {Foreman-Mackey}, {Fox},
  {Freij}, {Garg}, {Geda}, {Glattly}, {Gondhalekar}, {Gordon}, {Grant},
  {Greenfield}, {Groener}, {Guest}, {Gurovich}, {Handberg}, {Hart},
  {Hatfield-Dodds}, {Homeier}, {Hosseinzadeh}, {Jenness}, {Jones}, {Joseph},
  {Kalmbach}, {Karamehmetoglu}, {Ka{\l}uszy{\'n}ski}, {Kelley}, {Kern},
  {Kerzendorf}, {Koch}, {Kulumani}, {Lee}, {Ly}, {Ma}, {MacBride}, {Maljaars},
  {Muna}, {Murphy}, {Norman}, {O'Steen}, {Oman}, {Pacifici}, {Pascual},
  {Pascual-Granado}, {Patil}, {Perren}, {Pickering}, {Rastogi}, {Roulston},
  {Ryan}, {Rykoff}, {Sabater}, {Sakurikar}, {Salgado}, {Sanghi}, {Saunders},
  {Savchenko}, {Schwardt}, {Seifert-Eckert}, {Shih}, {Jain}, {Shukla}, {Sick},
  {Simpson}, {Singanamalla}, {Singer}, {Singhal}, {Sinha}, {Sip{\H{o}}cz},
  {Spitler}, {Stansby}, {Streicher}, {{\v{S}}umak}, {Swinbank}, {Taranu},
  {Tewary}, {Tremblay}, {de Val-Borro}, {Van Kooten}, {Vasovi{\'c}}, {Verma},
  {de Miranda Cardoso}, {Williams}, {Wilson}, {Winkel}, {Wood-Vasey}, {Xue},
  {Yoachim}, {Zhang}, {Zonca}, \& {Astropy Project Contributors}}]{astropy2022}
{Astropy Collaboration}, {Price-Whelan}, A.~M., {Lim}, P.~L., {et~al.} 2022,
  \apj, 935, 167

\bibitem[{{Astropy Collaboration} {et~al.}(2018){Astropy Collaboration},
  {Price-Whelan}, {Sip{\H{o}}cz}, {G{\"u}nther}, {Lim}, {Crawford}, {Conseil},
  {Shupe}, {Craig}, {Dencheva}, {Ginsburg}, {VanderPlas}, {Bradley},
  {P{\'e}rez-Su{\'a}rez}, {de Val-Borro}, {Aldcroft}, {Cruz}, {Robitaille},
  {Tollerud}, {Ardelean}, {Babej}, {Bach}, {Bachetti}, {Bakanov}, {Bamford},
  {Barentsen}, {Barmby}, {Baumbach}, {Berry}, {Biscani}, {Boquien}, {Bostroem},
  {Bouma}, {Brammer}, {Bray}, {Breytenbach}, {Buddelmeijer}, {Burke},
  {Calderone}, {Cano Rodr{\'\i}guez}, {Cara}, {Cardoso}, {Cheedella}, {Copin},
  {Corrales}, {Crichton}, {D'Avella}, {Deil}, {Depagne}, {Dietrich}, {Donath},
  {Droettboom}, {Earl}, {Erben}, {Fabbro}, {Ferreira}, {Finethy}, {Fox},
  {Garrison}, {Gibbons}, {Goldstein}, {Gommers}, {Greco}, {Greenfield},
  {Groener}, {Grollier}, {Hagen}, {Hirst}, {Homeier}, {Horton}, {Hosseinzadeh},
  {Hu}, {Hunkeler}, {Ivezi{\'c}}, {Jain}, {Jenness}, {Kanarek}, {Kendrew},
  {Kern}, {Kerzendorf}, {Khvalko}, {King}, {Kirkby}, {Kulkarni}, {Kumar},
  {Lee}, {Lenz}, {Littlefair}, {Ma}, {Macleod}, {Mastropietro}, {McCully},
  {Montagnac}, {Morris}, {Mueller}, {Mumford}, {Muna}, {Murphy}, {Nelson},
  {Nguyen}, {Ninan}, {N{\"o}the}, {Ogaz}, {Oh}, {Parejko}, {Parley}, {Pascual},
  {Patil}, {Patil}, {Plunkett}, {Prochaska}, {Rastogi}, {Reddy Janga},
  {Sabater}, {Sakurikar}, {Seifert}, {Sherbert}, {Sherwood-Taylor}, {Shih},
  {Sick}, {Silbiger}, {Singanamalla}, {Singer}, {Sladen}, {Sooley},
  {Sornarajah}, {Streicher}, {Teuben}, {Thomas}, {Tremblay}, {Turner},
  {Terr{\'o}n}, {van Kerkwijk}, {de la Vega}, {Watkins}, {Weaver}, {Whitmore},
  {Woillez}, {Zabalza}, \& {Astropy Contributors}}]{astropy2018}
{Astropy Collaboration}, {Price-Whelan}, A.~M., {Sip{\H{o}}cz}, B.~M., {et~al.}
  2018, \aj, 156, 123

\bibitem[{{Astropy Collaboration} {et~al.}(2013){Astropy Collaboration},
  {Robitaille}, {Tollerud}, {Greenfield}, {Droettboom}, {Bray}, {Aldcroft},
  {Davis}, {Ginsburg}, {Price-Whelan}, {Kerzendorf}, {Conley}, {Crighton},
  {Barbary}, {Muna}, {Ferguson}, {Grollier}, {Parikh}, {Nair}, {Unther},
  {Deil}, {Woillez}, {Conseil}, {Kramer}, {Turner}, {Singer}, {Fox}, {Weaver},
  {Zabalza}, {Edwards}, {Azalee Bostroem}, {Burke}, {Casey}, {Crawford},
  {Dencheva}, {Ely}, {Jenness}, {Labrie}, {Lim}, {Pierfederici}, {Pontzen},
  {Ptak}, {Refsdal}, {Servillat}, \& {Streicher}}]{astropy2013}
{Astropy Collaboration}, {Robitaille}, T.~P., {Tollerud}, E.~J., {et~al.} 2013,
  \aap, 558, A33

\bibitem[{{Bastian} \& {Lardo}(2018)}]{Bastian2018}
{Bastian}, N. \& {Lardo}, C. 2018, \araa, 56, 83

\bibitem[{{Bastian} {et~al.}(2013){Bastian}, {Schweizer}, {Goudfrooij},
  {Larsen}, \& {Kissler-Patig}}]{Bastian2013}
{Bastian}, N., {Schweizer}, F., {Goudfrooij}, P., {Larsen}, S.~S., \&
  {Kissler-Patig}, M. 2013, \mnras, 431, 1252

\bibitem[{{Beasley}(2020)}]{Beasley2020}
{Beasley}, M.~A. 2020, in Reviews in Frontiers of Modern Astrophysics; From
  Space Debris to Cosmology, 245--277

\bibitem[{{Bestenlehner} {et~al.}(2020){Bestenlehner}, {Crowther},
  {Caballero-Nieves}, {Schneider}, {Sim{\'o}n-D{\'\i}az}, {Brands}, {de Koter},
  {Gr{\"a}fener}, {Herrero}, {Langer}, {Lennon}, {Ma{\'\i}z Apell{\'a}niz},
  {Puls}, \& {Vink}}]{Bestenlehner2020}
{Bestenlehner}, J.~M., {Crowther}, P.~A., {Caballero-Nieves}, S.~M., {et~al.}
  2020, \mnras, 499, 1918

\bibitem[{{Bosch} {et~al.}(2009){Bosch}, {Terlevich}, \&
  {Terlevich}}]{Bosch2009}
{Bosch}, G., {Terlevich}, E., \& {Terlevich}, R. 2009, \aj, 137, 3437

\bibitem[{{Bouwens} {et~al.}(2021){Bouwens}, {Illingworth}, {van Dokkum},
  {Ribeiro}, {Oesch}, \& {Stefanon}}]{Bouwens2021}
{Bouwens}, R.~J., {Illingworth}, G.~D., {van Dokkum}, P.~G., {et~al.} 2021,
  \aj, 162, 255

\bibitem[{{Brandner} {et~al.}(2001){Brandner}, {Grebel}, {Barb{\'a}},
  {Walborn}, \& {Moneti}}]{Brandner2001}
{Brandner}, W., {Grebel}, E.~K., {Barb{\'a}}, R.~H., {Walborn}, N.~R., \&
  {Moneti}, A. 2001, \aj, 122, 858

\bibitem[{{Brands} {et~al.}(2022){Brands}, {de Koter}, {Bestenlehner},
  {Crowther}, {Sundqvist}, {Puls}, {Caballero-Nieves}, {Abdul-Masih},
  {Driessen}, {Garc{\'\i}a}, {Geen}, {Gr{\"a}fener}, {Hawcroft}, {Kaper},
  {Keszthelyi}, {Langer}, {Sana}, {Schneider}, {Shenar}, \&
  {Vink}}]{Brands2022}
{Brands}, S.~A., {de Koter}, A., {Bestenlehner}, J.~M., {et~al.} 2022, \aap,
  663, A36

\bibitem[{{Britavskiy} {et~al.}(2019){Britavskiy}, {Lennon}, {Patrick},
  {Evans}, {Herrero}, {Langer}, {van Loon}, {Clark}, {Schneider}, {Almeida},
  {Sana}, {de Koter}, \& {Taylor}}]{Britavskiy2019}
{Britavskiy}, N., {Lennon}, D.~J., {Patrick}, L.~R., {et~al.} 2019, \aap, 624,
  A128

\bibitem[{{Brodie} \& {Strader}(2006)}]{Brodie2006}
{Brodie}, J.~P. \& {Strader}, J. 2006, \araa, 44, 193

\bibitem[{{Brunet} {et~al.}(1975){Brunet}, {Imbert}, {Martin}, {Mianes},
  {Pr{\'e}vot}, {Rebeirot}, \& {Rousseau}}]{Brunet1975}
{Brunet}, J.~P., {Imbert}, M., {Martin}, N., {et~al.} 1975, \aaps, 21, 109

\bibitem[{{Cardelli} {et~al.}(1989){Cardelli}, {Clayton}, \&
  {Mathis}}]{Cardelli1989}
{Cardelli}, J.~A., {Clayton}, G.~C., \& {Mathis}, J.~S. 1989, \apj, 345, 245

\bibitem[{{Carlson} {et~al.}(2012){Carlson}, {Sewi{\l}o}, {Meixner}, {Romita},
  \& {Lawton}}]{Carlson2012}
{Carlson}, L.~R., {Sewi{\l}o}, M., {Meixner}, M., {Romita}, K.~A., \& {Lawton},
  B. 2012, \aap, 542, A66

\bibitem[{{Cheng} {et~al.}(2021){Cheng}, {Wang}, \& {Lim}}]{Cheng2021}
{Cheng}, Y., {Wang}, Q.~D., \& {Lim}, S. 2021, \mnras, 504, 1627

\bibitem[{{Choi} {et~al.}(2016){Choi}, {Dotter}, {Conroy}, {Cantiello},
  {Paxton}, \& {Johnson}}]{Choi2016}
{Choi}, J., {Dotter}, A., {Conroy}, C., {et~al.} 2016, \apj, 823, 102

\bibitem[{{Choudhury} {et~al.}(2021){Choudhury}, {de Grijs}, {Bekki}, {Cioni},
  {Ivanov}, {van Loon}, {Miller}, {Niederhofer}, {Oliveira}, {Ripepi}, {Sun},
  \& {Subramanian}}]{Choudhury2021}
{Choudhury}, S., {de Grijs}, R., {Bekki}, K., {et~al.} 2021, \mnras, 507, 4752

\bibitem[{{Cignoni} {et~al.}(2016){Cignoni}, {Sabbi}, {van der Marel},
  {Lennon}, {Tosi}, {Grebel}, {Gallagher}, {Aloisi}, {de Marchi}, {Gouliermis},
  {Larsen}, {Panagia}, \& {Smith}}]{Cignoni2016}
{Cignoni}, M., {Sabbi}, E., {van der Marel}, R.~P., {et~al.} 2016, \apj, 833,
  154

\bibitem[{{Cignoni} {et~al.}(2015){Cignoni}, {Sabbi}, {van der Marel}, {Tosi},
  {Zaritsky}, {Anderson}, {Lennon}, {Aloisi}, {de Marchi}, {Gouliermis},
  {Grebel}, {Smith}, \& {Zeidler}}]{Cignoni2015}
{Cignoni}, M., {Sabbi}, E., {van der Marel}, R.~P., {et~al.} 2015, \apj, 811,
  76

\bibitem[{{Crowther} {et~al.}(2022){Crowther}, {Broos}, {Townsley}, {Pollock},
  {Tehrani}, \& {Gagn{\'e}}}]{Crowther2022}
{Crowther}, P.~A., {Broos}, P.~S., {Townsley}, L.~K., {et~al.} 2022, \mnras,
  515, 4130

\bibitem[{{Crowther} {et~al.}(2016){Crowther}, {Caballero-Nieves}, {Bostroem},
  {Ma{\'\i}z Apell{\'a}niz}, {Schneider}, {Walborn}, {Angus}, {Brott},
  {Bonanos}, {de Koter}, {de Mink}, {Evans}, {Gr{\"a}fener}, {Herrero},
  {Howarth}, {Langer}, {Lennon}, {Puls}, {Sana}, \& {Vink}}]{Crowther2016}
{Crowther}, P.~A., {Caballero-Nieves}, S.~M., {Bostroem}, K.~A., {et~al.} 2016,
  \mnras, 458, 624

\bibitem[{{Crowther} {et~al.}(2010){Crowther}, {Schnurr}, {Hirschi}, {Yusof},
  {Parker}, {Goodwin}, \& {Kassim}}]{Crowther2010}
{Crowther}, P.~A., {Schnurr}, O., {Hirschi}, R., {et~al.} 2010, \mnras, 408,
  731

\bibitem[{{De Marchi} {et~al.}(2013){De Marchi}, {Beccari}, \&
  {Panagia}}]{DeMarchi2013}
{De Marchi}, G., {Beccari}, G., \& {Panagia}, N. 2013, \apj, 775, 68

\bibitem[{{De Marchi} {et~al.}(2023){De Marchi}, {Giardino}, {Biazzo},
  {Panagia}, {Sabbi}, {Beck}, {Robberto}, {Zeidler}, {Jones}, {Mexiner},
  {Fahrion}, {Habel}, {Nally}, {Hirschauer}, {Soderblom}, {Nayak}, {Lenki},
  {Rogers}, {Brandl}, \& {Keyes}}]{DeMarchi2023}
{De Marchi}, G., {Giardino}, G., {Biazzo}, K., {et~al.} 2023, submitted to
  Nature Astronomy

\bibitem[{{De Marchi} {et~al.}(2017){De Marchi}, {Panagia}, \&
  {Beccari}}]{DeMarchi2017}
{De Marchi}, G., {Panagia}, N., \& {Beccari}, G. 2017, \apj, 846, 110

\bibitem[{{De Marchi} {et~al.}(2010){De Marchi}, {Panagia}, \&
  {Romaniello}}]{DeMarchi2010}
{De Marchi}, G., {Panagia}, N., \& {Romaniello}, M. 2010, \apj, 715, 1

\bibitem[{{De Marchi} {et~al.}(2011{\natexlab{a}}){De Marchi}, {Panagia}, \&
  {Sabbi}}]{DeMarchi2011}
{De Marchi}, G., {Panagia}, N., \& {Sabbi}, E. 2011{\natexlab{a}}, \apj, 740,
  10

\bibitem[{{De Marchi} {et~al.}(2011{\natexlab{b}}){De Marchi}, {Paresce},
  {Panagia}, {Beccari}, {Spezzi}, {Sirianni}, {Andersen}, {Mutchler}, {Balick},
  {Dopita}, {Frogel}, {Whitmore}, {Bond}, {Calzetti}, {Carollo}, {Disney},
  {Hall}, {Holtzman}, {Kimble}, {McCarthy}, {O'Connell}, {Saha}, {Silk},
  {Trauger}, {Walker}, {Windhorst}, \& {Young}}]{DeMarchi2011a}
{De Marchi}, G., {Paresce}, F., {Panagia}, N., {et~al.} 2011{\natexlab{b}},
  \apj, 739, 27

\bibitem[{{Dom{\'\i}nguez} {et~al.}(2023){Dom{\'\i}nguez}, {Pellegrini},
  {Klessen}, \& {Rahner}}]{Dominguez2023}
{Dom{\'\i}nguez}, R., {Pellegrini}, E.~W., {Klessen}, R.~S., \& {Rahner}, D.
  2023, \mnras, 520, 5600

\bibitem[{{Dotter}(2016)}]{Dotter2016}
{Dotter}, A. 2016, \apjs, 222, 8

\bibitem[{{Evans} {et~al.}(2015){Evans}, {Kennedy}, {Dufton}, {Howarth},
  {Walborn}, {Markova}, {Clark}, {de Mink}, {de Koter}, {Dunstall},
  {H{\'e}nault-Brunet}, {Ma{\'\i}z Apell{\'a}niz}, {McEvoy}, {Sana},
  {Sim{\'o}n-D{\'\i}az}, {Taylor}, \& {Vink}}]{Evans2015}
{Evans}, C.~J., {Kennedy}, M.~B., {Dufton}, P.~L., {et~al.} 2015, \aap, 574,
  A13

\bibitem[{{Evans} {et~al.}(2011){Evans}, {Taylor}, {H{\'e}nault-Brunet},
  {Sana}, {de Koter}, {Sim{\'o}n-D{\'\i}az}, {Carraro}, {Bagnoli}, {Bastian},
  {Bestenlehner}, {Bonanos}, {Bressert}, {Brott}, {Campbell}, {Cantiello},
  {Clark}, {Costa}, {Crowther}, {de Mink}, {Doran}, {Dufton}, {Dunstall},
  {Friedrich}, {Garcia}, {Gieles}, {Gr{\"a}fener}, {Herrero}, {Howarth},
  {Izzard}, {Langer}, {Lennon}, {Ma{\'\i}z Apell{\'a}niz}, {Markova},
  {Najarro}, {Puls}, {Ramirez}, {Sab{\'\i}n-Sanjuli{\'a}n}, {Smartt}, {Stroud},
  {van Loon}, {Vink}, \& {Walborn}}]{Evans2011}
{Evans}, C.~J., {Taylor}, W.~D., {H{\'e}nault-Brunet}, V., {et~al.} 2011, \aap,
  530, A108

\bibitem[{{Fahrion} \& {De Marchi}(2023)}]{FahrionDeMarchi2023}
{Fahrion}, K. \& {De Marchi}, G. 2023, \aap, 671, L14

\bibitem[{{Fitzpatrick} \& {Savage}(1984)}]{FitzpatrickSavage1984}
{Fitzpatrick}, E.~L. \& {Savage}, B.~D. 1984, \apj, 279, 578

\bibitem[{{Forbes} {et~al.}(2018){Forbes}, {Bastian}, {Gieles}, {Crain},
  {Kruijssen}, {Larsen}, {Ploeckinger}, {Agertz}, {Trenti}, {Ferguson},
  {Pfeffer}, \& {Gnedin}}]{Forbes2018}
{Forbes}, D.~A., {Bastian}, N., {Gieles}, M., {et~al.} 2018, Proceedings of the
  Royal Society of London Series A, 474, 20170616

\bibitem[{{Grebel} \& {Chu}(2000)}]{GrebelChu2000}
{Grebel}, E.~K. \& {Chu}, Y.-H. 2000, \aj, 119, 787

\bibitem[{{Gruendl} \& {Chu}(2009)}]{GruendlChu2009}
{Gruendl}, R.~A. \& {Chu}, Y.-H. 2009, \apjs, 184, 172

\bibitem[{{Guszejnov} {et~al.}(2022){Guszejnov}, {Markey}, {Offner},
  {Grudi{\'c}}, {Faucher-Gigu{\`e}re}, {Rosen}, \& {Hopkins}}]{Guszejnov2022}
{Guszejnov}, D., {Markey}, C., {Offner}, S. S.~R., {et~al.} 2022, \mnras, 515,
  167

\bibitem[{{Howard} {et~al.}(2018){Howard}, {Pudritz}, \& {Harris}}]{Howard2018}
{Howard}, C.~S., {Pudritz}, R.~E., \& {Harris}, W.~E. 2018, Nature Astronomy,
  2, 725

\bibitem[{{Hunter} {et~al.}(1995{\natexlab{a}}){Hunter}, {Shaya}, {Holtzman},
  {Light}, {O'Neil}, \& {Lynds}}]{Hunter1995}
{Hunter}, D.~A., {Shaya}, E.~J., {Holtzman}, J.~A., {et~al.}
  1995{\natexlab{a}}, \apj, 448, 179

\bibitem[{{Hunter} {et~al.}(1995{\natexlab{b}}){Hunter}, {Shaya}, {Scowen},
  {Hester}, {Groth}, {Lynds}, \& {O'Neil}}]{Hunter1995a}
{Hunter}, D.~A., {Shaya}, E.~J., {Scowen}, P., {et~al.} 1995{\natexlab{b}},
  \apj, 444, 758

\bibitem[{{Hyland} {et~al.}(1992){Hyland}, {Straw}, {Jones}, \&
  {Gatley}}]{Hyland1992}
{Hyland}, A.~R., {Straw}, S., {Jones}, T.~J., \& {Gatley}, I. 1992, \mnras,
  257, 391

\bibitem[{{Isserstedt}(1975)}]{Isserstedt1975}
{Isserstedt}, J. 1975, \aap, 41, 175

\bibitem[{{Jones} {et~al.}(2023){Jones}, {Nally}, {Habel}, {Lenki{\'c}},
  {Fahrion}, {Hirschauer}, {Chu}, {Meixner}, {De Marchi}, {Nayak}, {Robberto},
  {Sabbi}, {Zeidler}, {Alves de Oliveira}, {Beck}, {Biazzo}, {Brandl},
  {Giardino}, {Jerabkova}, {Keyes}, {Muzerolle}, {Panagia}, {Pontoppidan},
  {Rogers}, {Sargent}, \& {Soderblom}}]{Jones2023}
{Jones}, O.~C., {Nally}, C., {Habel}, N., {et~al.} 2023, Nature Astronomy, 7,
  694

\bibitem[{{Jones} {et~al.}(2022){Jones}, {Reiter}, {Sanchez-Janssen}, {Evans},
  {Robertson}, {Meixner}, \& {Ochsendorf}}]{Jones2022}
{Jones}, O.~C., {Reiter}, M., {Sanchez-Janssen}, R., {et~al.} 2022, \mnras,
  517, 1518

\bibitem[{{Kalari} {et~al.}(2018){Kalari}, {Rubio}, {Elmegreen}, {Guzm{\'a}n},
  {Zinnecker}, \& {Herrera}}]{Kalari2018}
{Kalari}, V.~M., {Rubio}, M., {Elmegreen}, B.~G., {et~al.} 2018, \apj, 852, 71

\bibitem[{{Krause} {et~al.}(2020){Krause}, {Offner}, {Charbonnel}, {Gieles},
  {Klessen}, {V{\'a}zquez-Semadeni}, {Ballesteros-Paredes}, {Girichidis},
  {Kruijssen}, {Ward}, \& {Zinnecker}}]{Krause2020}
{Krause}, M. G.~H., {Offner}, S. S.~R., {Charbonnel}, C., {et~al.} 2020, \ssr,
  216, 64

\bibitem[{{Krumholz} {et~al.}(2019){Krumholz}, {McKee}, \&
  {Bland-Hawthorn}}]{Krumholz2019}
{Krumholz}, M.~R., {McKee}, C.~F., \& {Bland-Hawthorn}, J. 2019, \araa, 57, 227

\bibitem[{{Ksoll} {et~al.}(2018){Ksoll}, {Gouliermis}, {Klessen}, {Grebel},
  {Sabbi}, {Anderson}, {Lennon}, {Cignoni}, {de Marchi}, {Smith}, {Tosi}, \&
  {van der Marel}}]{Ksoll2018}
{Ksoll}, V.~F., {Gouliermis}, D.~A., {Klessen}, R.~S., {et~al.} 2018, \mnras,
  479, 2389

\bibitem[{{Lah{\'e}n} {et~al.}(2020){Lah{\'e}n}, {Naab}, {Johansson},
  {Elmegreen}, {Hu}, {Walch}, {Steinwandel}, \& {Moster}}]{Lahen2020}
{Lah{\'e}n}, N., {Naab}, T., {Johansson}, P.~H., {et~al.} 2020, \apj, 891, 2

\bibitem[{{Lenki{\'c}} {et~al.}(2023){Lenki{\'c}}, {Nally}, {Jones}, {Boyer},
  {Kavanagh}, {Habel}, {Nayak}, {Hirschauer}, {Meixner}, {Sargent}, \&
  {Temim}}]{Lenkic2023}
{Lenki{\'c}}, L., {Nally}, C., {Jones}, O.~C., {et~al.} 2023, arXiv e-prints,
  arXiv:2307.15704

\bibitem[{{Lortet} \& {Testor}(1991)}]{Lortet1991}
{Lortet}, M.~C. \& {Testor}, G. 1991, \aaps, 89, 185

\bibitem[{{Ma{\'\i}z Apell{\'a}niz} \& {Barb{\'a}}(2018)}]{MaizApellaniz2018}
{Ma{\'\i}z Apell{\'a}niz}, J. \& {Barb{\'a}}, R.~H. 2018, \aap, 613, A9

\bibitem[{{Ma{\'\i}z Apell{\'a}niz} {et~al.}(2014){Ma{\'\i}z Apell{\'a}niz},
  {Evans}, {Barb{\'a}}, {Gr{\"a}fener}, {Bestenlehner}, {Crowther},
  {Garc{\'\i}a}, {Herrero}, {Sana}, {Sim{\'o}n-D{\'\i}az}, {Taylor}, {van
  Loon}, {Vink}, \& {Walborn}}]{MaizApellaniz2014}
{Ma{\'\i}z Apell{\'a}niz}, J., {Evans}, C.~J., {Barb{\'a}}, R.~H., {et~al.}
  2014, \aap, 564, A63

\bibitem[{{Ma{\'\i}z Apell{\'a}niz} {et~al.}(2020){Ma{\'\i}z Apell{\'a}niz},
  {Pantaleoni Gonz{\'a}lez}, {Barb{\'a}}, {Garc{\'\i}a-Lario}, \&
  {Nogueras-Lara}}]{MaizApellaniz2020}
{Ma{\'\i}z Apell{\'a}niz}, J., {Pantaleoni Gonz{\'a}lez}, M., {Barb{\'a}},
  R.~H., {Garc{\'\i}a-Lario}, P., \& {Nogueras-Lara}, F. 2020, \mnras, 496,
  4951

\bibitem[{{Meixner} {et~al.}(2006){Meixner}, {Gordon}, {Indebetouw}, {Hora},
  {Whitney}, {Blum}, {Reach}, {Bernard}, {Meade}, {Babler}, {Engelbracht},
  {For}, {Misselt}, {Vijh}, {Leitherer}, {Cohen}, {Churchwell}, {Boulanger},
  {Frogel}, {Fukui}, {Gallagher}, {Gorjian}, {Harris}, {Kelly}, {Kawamura},
  {Kim}, {Latter}, {Madden}, {Markwick-Kemper}, {Mizuno}, {Mizuno}, {Mould},
  {Nota}, {Oey}, {Olsen}, {Onishi}, {Paladini}, {Panagia}, {Perez-Gonzalez},
  {Shibai}, {Sato}, {Smith}, {Staveley-Smith}, {Tielens}, {Ueta}, {van Dyk},
  {Volk}, {Werner}, \& {Zaritsky}}]{Meixner2006}
{Meixner}, M., {Gordon}, K.~D., {Indebetouw}, R., {et~al.} 2006, \aj, 132, 2268

\bibitem[{{Meixner} {et~al.}(2013){Meixner}, {Panuzzo}, {Roman-Duval},
  {Engelbracht}, {Babler}, {Seale}, {Hony}, {Montiel}, {Sauvage}, {Gordon},
  {Misselt}, {Okumura}, {Chanial}, {Beck}, {Bernard}, {Bolatto}, {Bot},
  {Boyer}, {Carlson}, {Clayton}, {Chen}, {Cormier}, {Fukui}, {Galametz},
  {Galliano}, {Hora}, {Hughes}, {Indebetouw}, {Israel}, {Kawamura}, {Kemper},
  {Kim}, {Kwon}, {Lebouteiller}, {Li}, {Long}, {Madden}, {Matsuura}, {Muller},
  {Oliveira}, {Onishi}, {Otsuka}, {Paradis}, {Poglitsch}, {Reach},
  {Robitaille}, {Rubio}, {Sargent}, {Sewi{\l}o}, {Skibba}, {Smith},
  {Srinivasan}, {Tielens}, {van Loon}, \& {Whitney}}]{Meixner2013}
{Meixner}, M., {Panuzzo}, P., {Roman-Duval}, J., {et~al.} 2013, \aj, 146, 62

\bibitem[{{Nally} {et~al.}(2023){Nally}, {Jones}, {Lenki{\'c}}, {Habel},
  {Hirschauer}, {Meixner}, {Kavanagh}, {Boyer}, {Ferguson}, {Sargent}, {Nayak},
  \& {Temim}}]{Nally2023}
{Nally}, C., {Jones}, O.~C., {Lenki{\'c}}, L., {et~al.} 2023, arXiv e-prints,
  arXiv:2309.13521

\bibitem[{{Nayak} {et~al.}(2023){Nayak}, {Green}, {Hirschauer}, {Indebetouw},
  {Meixner}, {Wong}, {Chevance}, {De Marchi}, {Lebouteiller}, {Lee}, {Looney},
  {Madden}, {Roman-Duval}, {Fukui}, {Hacar}, {Jameson}, {Kalari}, {Oudshoorn},
  {Rubio}, \& {Sabbi}}]{Nayak2023}
{Nayak}, O., {Green}, A., {Hirschauer}, A.~S., {et~al.} 2023, \apj, 944, 26

\bibitem[{{Panagia}(2005)}]{Panagia2005}
{Panagia}, N. 2005, in IAU Colloq. 192: Cosmic Explosions, On the 10th
  Anniversary of SN1993J, ed. J.-M. {Marcaide} \& K.~W. {Weiler}, Vol.~99, 585

\bibitem[{{Panagia} {et~al.}(1991){Panagia}, {Gilmozzi}, {Macchetto}, {Adorf},
  \& {Kirshner}}]{Panagia1991}
{Panagia}, N., {Gilmozzi}, R., {Macchetto}, F., {Adorf}, H.~M., \& {Kirshner},
  R.~P. 1991, \apjl, 380, L23

\bibitem[{{Portegies Zwart} {et~al.}(2010){Portegies Zwart}, {McMillan}, \&
  {Gieles}}]{Portegies2010}
{Portegies Zwart}, S.~F., {McMillan}, S. L.~W., \& {Gieles}, M. 2010, \araa,
  48, 431

\bibitem[{{Robitaille} {et~al.}(2007){Robitaille}, {Whitney}, {Indebetouw}, \&
  {Wood}}]{Robitaille2007}
{Robitaille}, T.~P., {Whitney}, B.~A., {Indebetouw}, R., \& {Wood}, K. 2007,
  \apjs, 169, 328

\bibitem[{{Rubio} {et~al.}(1998){Rubio}, {Barb{\'a}}, {Walborn}, {Probst},
  {Garc{\'\i}a}, \& {Roth}}]{Rubio1998}
{Rubio}, M., {Barb{\'a}}, R.~H., {Walborn}, N.~R., {et~al.} 1998, \aj, 116,
  1708

\bibitem[{{Rubio} {et~al.}(1992){Rubio}, {Roth}, \& {Garcia}}]{Rubio1992}
{Rubio}, M., {Roth}, M., \& {Garcia}, J. 1992, \aap, 261, L29

\bibitem[{{Sabbi} {et~al.}(2013){Sabbi}, {Anderson}, {Lennon}, {van der Marel},
  {Aloisi}, {Boyer}, {Cignoni}, {de Marchi}, {de Mink}, {Evans}, {Gallagher},
  {Gordon}, {Gouliermis}, {Grebel}, {Koekemoer}, {Larsen}, {Panagia}, {Ryon},
  {Smith}, {Tosi}, \& {Zaritsky}}]{Sabbi2013}
{Sabbi}, E., {Anderson}, J., {Lennon}, D.~J., {et~al.} 2013, \aj, 146, 53

\bibitem[{{Sabbi} {et~al.}(2016){Sabbi}, {Lennon}, {Anderson}, {Cignoni}, {van
  der Marel}, {Zaritsky}, {De Marchi}, {Panagia}, {Gouliermis}, {Grebel},
  {Gallagher}, {Smith}, {Sana}, {Aloisi}, {Tosi}, {Evans}, {Arab}, {Boyer}, {de
  Mink}, {Gordon}, {Koekemoer}, {Larsen}, {Ryon}, \& {Zeidler}}]{Sabbi2016}
{Sabbi}, E., {Lennon}, D.~J., {Anderson}, J., {et~al.} 2016, \apjs, 222, 11

\bibitem[{{Sabbi} {et~al.}(2012){Sabbi}, {Lennon}, {Gieles}, {de Mink},
  {Walborn}, {Anderson}, {Bellini}, {Panagia}, {van der Marel}, \& {Ma{\'\i}z
  Apell{\'a}niz}}]{Sabbi2012}
{Sabbi}, E., {Lennon}, D.~J., {Gieles}, M., {et~al.} 2012, \apjl, 754, L37

\bibitem[{{Schneider} {et~al.}(2018){Schneider}, {Ram{\'\i}rez-Agudelo},
  {Tramper}, {Bestenlehner}, {Castro}, {Sana}, {Evans},
  {Sab{\'\i}n-Sanjuli{\'a}n}, {Sim{\'o}n-D{\'\i}az}, {Langer}, {Fossati},
  {Gr{\"a}fener}, {Crowther}, {de Mink}, {de Koter}, {Gieles}, {Herrero},
  {Izzard}, {Kalari}, {Klessen}, {Lennon}, {Mahy}, {Ma{\'\i}z Apell{\'a}niz},
  {Markova}, {van Loon}, {Vink}, \& {Walborn}}]{Schneider2018}
{Schneider}, F.~R.~N., {Ram{\'\i}rez-Agudelo}, O.~H., {Tramper}, F., {et~al.}
  2018, \aap, 618, A73

\bibitem[{{Scowen} {et~al.}(1998){Scowen}, {Hester}, {Sankrit}, {Gallagher},
  {Ballester}, {Burrows}, {Clarke}, {Crisp}, {Evans}, {Griffiths}, {Hoessel},
  {Holtzman}, {Krist}, {Mould}, {Stapelfeldt}, {Trauger}, {Watson}, \&
  {Westphal}}]{Scowen1998}
{Scowen}, P.~A., {Hester}, J.~J., {Sankrit}, R., {et~al.} 1998, \aj, 116, 163

\bibitem[{{Selman} {et~al.}(1999){Selman}, {Melnick}, {Bosch}, \&
  {Terlevich}}]{Selman1999}
{Selman}, F., {Melnick}, J., {Bosch}, G., \& {Terlevich}, R. 1999, \aap, 347,
  532

\bibitem[{{Tsilia} {et~al.}(2023){Tsilia}, {De Marchi}, \&
  {Panagia}}]{Tsilia2023}
{Tsilia}, S., {De Marchi}, G., \& {Panagia}, N. 2023, \aap, 675, A203

\bibitem[{{Vlasblom} \& {De Marchi}(2023)}]{Vlasblom2023}
{Vlasblom}, M. \& {De Marchi}, G. 2023, \aap, 675, A204

\bibitem[{{Walborn}(1991)}]{Walborn1991}
{Walborn}, N.~R. 1991, in The Magellanic Clouds, ed. R.~{Haynes} \& D.~{Milne},
  Vol. 148, 145

\bibitem[{{Walborn} {et~al.}(1999){Walborn}, {Barb{\'a}}, {Brandner}, {Rubio},
  {Grebel}, \& {Probst}}]{Walborn1999}
{Walborn}, N.~R., {Barb{\'a}}, R.~H., {Brandner}, W., {et~al.} 1999, \aj, 117,
  225

\bibitem[{{Walborn} {et~al.}(2013){Walborn}, {Barb{\'a}}, \&
  {Sewi{\l}o}}]{Walborn2013}
{Walborn}, N.~R., {Barb{\'a}}, R.~H., \& {Sewi{\l}o}, M.~M. 2013, \aj, 145, 98

\bibitem[{{Walborn} \& {Blades}(1987)}]{Walborn1987}
{Walborn}, N.~R. \& {Blades}, J.~C. 1987, \apjl, 323, L65

\bibitem[{{Walborn} {et~al.}(2002){Walborn}, {Ma{\'\i}z-Apell{\'a}niz}, \&
  {Barb{\'a}}}]{Walborn2002}
{Walborn}, N.~R., {Ma{\'\i}z-Apell{\'a}niz}, J., \& {Barb{\'a}}, R.~H. 2002,
  \aj, 124, 1601

\bibitem[{{Whitmore} {et~al.}(2010){Whitmore}, {Chandar}, {Schweizer},
  {Rothberg}, {Leitherer}, {Rieke}, {Rieke}, {Blair}, {Mengel}, \&
  {Alonso-Herrero}}]{Whitmore2010}
{Whitmore}, B.~C., {Chandar}, R., {Schweizer}, F., {et~al.} 2010, \aj, 140, 75

\bibitem[{{Whitney} {et~al.}(2008){Whitney}, {Sewilo}, {Indebetouw},
  {Robitaille}, {Meixner}, {Gordon}, {Meade}, {Babler}, {Harris}, {Hora},
  {Bracker}, {Povich}, {Churchwell}, {Engelbracht}, {For}, {Block}, {Misselt},
  {Vijh}, {Leitherer}, {Kawamura}, {Blum}, {Cohen}, {Fukui}, {Mizuno},
  {Mizuno}, {Srinivasan}, {Tielens}, {Volk}, {Bernard}, {Boulanger}, {Frogel},
  {Gallagher}, {Gorjian}, {Kelly}, {Latter}, {Madden}, {Kemper}, {Mould},
  {Nota}, {Oey}, {Olsen}, {Onishi}, {Paladini}, {Panagia}, {Perez-Gonzalez},
  {Reach}, {Shibai}, {Sato}, {Smith}, {Staveley-Smith}, {Ueta}, {Van Dyk},
  {Werner}, {Wolff}, \& {Zaritsky}}]{Whitney2008}
{Whitney}, B.~A., {Sewilo}, M., {Indebetouw}, R., {et~al.} 2008, \aj, 136, 18

\bibitem[{{Wong} {et~al.}(2022){Wong}, {Oudshoorn}, {Sofovich}, {Green},
  {Shah}, {Indebetouw}, {Meixner}, {Hacar}, {Nayak}, {Tokuda}, {Bolatto},
  {Chevance}, {De Marchi}, {Fukui}, {Hirschauer}, {Jameson}, {Kalari},
  {Lebouteiller}, {Looney}, {Madden}, {Onishi}, {Roman-Duval}, {Rubio}, \&
  {Tielens}}]{Wong2022}
{Wong}, T., {Oudshoorn}, L., {Sofovich}, E., {et~al.} 2022, \apj, 932, 47

\end{thebibliography}

\begin{appendix}

\onecolumn
\section{Comparing aperture and PSF photometry}
\label{app:aperture_corrections}
We list in Table \ref{tab:aperture_corrections} the derived aperture corrections for our default choice of an 2.5 aperture with a background annulus between 4 and 5.5 pixels, as well as other choices. These corrections were determined by calculating the offset between aperture photometry and PSF photometry for all unsaturated stars brighter than 20 mag in the respective filter. 

\begin{table*}
    \caption{Aperture corrections in magnitudes for different choices of aperture radius.}
    \centering
    \begin{tabular}{c c c c c c c} \hline\hline
       Field & Filter  & C$_\text{ap, 2 pix}$ & C$_\text{ap, 2.5 pix}$ & C$_\text{ap, 3 pix}$ & C$_\text{ap, 3.5 pix}$ &  C$_\text{ap, 4 pix}$   \\
        & & (mag) & (mag) & (mag) & (mag) & (mag) \\ \hline
        Centre & F090W & 0.386 $\pm$ 0.026 & 0.297 $\pm$ 0.025 & 0.244 $\pm$ 0.025 & 0.204 $\pm$ 0.026 & 0.171 $\pm$ 0.028 \\
       Centre & F187N & 0.551 $\pm$ 0.050 & 0.432 $\pm$ 0.043 & 0.359 $\pm$ 0.054 & 0.288 $\pm$ 0.077 & 0.228 $\pm$ 0.101 \\
       Centre & F200W & 0.562 $\pm$ 0.021 & 0.447 $\pm$ 0.021 & 0.383 $\pm$ 0.022 & 0.319 $\pm$ 0.026 & 0.258 $\pm$ 0.031 \\
       Centre & F335M & 0.444 $\pm$ 0.057 & 0.326 $\pm$ 0.059 & 0.243 $\pm$ 0.082 & 0.175 $\pm$ 0.118 & 0.127 $\pm$ 0.166 \\
      Centre &  F444W & 0.580 $\pm$ 0.046 & 0.440 $\pm$ 0.055 & 0.366 $\pm$ 0.081 & 0.298 $\pm$ 0.121 & 0.226 $\pm$ 0.167 \\
       Centre &  F470N & 0.647 $\pm$ 0.057 & 0.487 $\pm$ 0.065 & 0.409 $\pm$ 0.088 & 0.349 $\pm$ 0.119 & 0.284 $\pm$ 0.154 \\ \hline
       Northern parallel & F182M & 0.523 $\pm$ 0.020 & 0.421 $\pm$ 0.018 & 0.357 $\pm$ 0.016 & 0.289 $\pm$ 0.017 & 0.236 $\pm$ 0.019 \\
       Northern parallel & F187N & 0.531 $\pm$ 0.022 & 0.427 $\pm$ 0.021 & 0.362 $\pm$ 0.021 & 0.294 $\pm$ 0.022 & 0.235 $\pm$ 0.026 \\
       Northern parallel & F405N & 0.537 $\pm$ 0.044 & 0.411 $\pm$ 0.057 & 0.342 $\pm$ 0.077 & 0.276 $\pm$ 0.104 & 0.210 $\pm$ 0.125 \\
       Northern parallel & F430M & 0.568 $\pm$ 0.022 & 0.438 $\pm$ 0.024 & 0.369 $\pm$ 0.028 & 0.310 $\pm$ 0.032 & 0.246 $\pm$ 0.040 \\ \hline
       Southern parallel & F187N & 0.532 $\pm$ 0.030 & 0.425 $\pm$ 0.028 & 0.358 $\pm$ 0.030 & 0.289 $\pm$ 0.032 & 0.231 $\pm$ 0.037 \\
       Southern parallel & F200W & 0.563 $\pm$ 0.021 & 0.445 $\pm$ 0.020 & 0.383 $\pm$ 0.018 & 0.321 $\pm$ 0.017 & 0.262 $\pm$ 0.018 \\
       Southern parallel & F212N & 0.610 $\pm$ 0.031 & 0.480 $\pm$ 0.029 & 0.411 $\pm$ 0.034 & 0.352 $\pm$ 0.037 & 0.287 $\pm$ 0.036 \\
       Southern parallel & F405N & 0.538 $\pm$ 0.052 & 0.405 $\pm$ 0.071 & 0.328 $\pm$ 0.097 & 0.259 $\pm$ 0.119 & 0.189 $\pm$ 0.156 \\
       Southern parallel & F444W & 0.580 $\pm$ 0.024 & 0.445 $\pm$ 0.025 & 0.372 $\pm$ 0.030 & 0.310 $\pm$ 0.040 & 0.244 $\pm$ 0.052 \\
       Southern parallel & F470N & 0.640 $\pm$ 0.065 & 0.488 $\pm$ 0.075 & 0.411 $\pm$ 0.096 & 0.347 $\pm$ 0.132 & 0.284 $\pm$ 0.165 \\ \hline
       
    \end{tabular}
    \tablefoot{These values were calculated using stars brighter than 20 mag. The chosen values for the inner and outer radius of the sky annulus are (3.5, 5), (4, 5.5), (4.5, 6), (5, 6.5), and (5.5, 7) pixels, respectively.}
    \label{tab:aperture_corrections}
\end{table*}

\end{appendix}

\end{document}